\documentclass[journal]{IEEEtran}
\usepackage[utf8]{inputenc}
\usepackage{color}
\usepackage{xcolor}
\usepackage{array}
\usepackage{verbatim}
\usepackage{float}
\usepackage{amsmath}
\usepackage{amsthm}
\usepackage{amssymb}
\usepackage{graphicx}
\usepackage{longtable}
\usepackage{multirow}
\usepackage{booktabs}
\usepackage[unicode=true,
bookmarks=false,
breaklinks=false,pdfborder={0 0 1},colorlinks=false]
{hyperref}
\hypersetup{
	colorlinks,bookmarksopen,bookmarksnumbered,citecolor=blue,urlcolor=blue}
\usepackage{cite}

\usepackage{lipsum}
\usepackage{mathtools}
\usepackage{cuted}
\providecommand{\tabularnewline}{\\}
\usepackage{algorithmic}
\usepackage{longtable}

\floatstyle{ruled}
\newfloat{algorithm}{tbp}{loa}
\providecommand{\algorithmname}{Algorithm}
\floatname{algorithm}{\protect\algorithmname}

\makeatletter
\let\oldforeign@language\foreign@language
\DeclareRobustCommand{\foreign@language}[1]{%
	\lowercase{\oldforeign@language{#1}}}

\let\oldforeign@language\foreign@language
\DeclareRobustCommand{\foreign@language}[1]{%
	\lowercase{\oldforeign@language{#1}}}

\ifCLASSINFOpdf
\else
\fi

\@ifundefined{showcaptionsetup}{}{%
	\PassOptionsToPackage{caption=false}{subfig}}
\usepackage{subfig}

\usepackage{balance}

\ifCLASSINFOpdf
\else
\fi

\newtheorem{rem}{Remark}

\newtheorem{assum}{Assumption}
\ifCLASSINFOpdf
\else
\fi

\def\ps@IEEEtitlepagestyle{%
	\def\@oddhead{\parbox[t][\height][t]{\textwidth}{\centering \scriptsize
			Personal use of this material is permitted. Permission from the author(s) and/or copyright holder(s), must be obtained for all other uses. Please contact us and provide details if you believe this document breaches copyrights.\\
			\noindent\makebox[\linewidth]{}
		}\hfil\hbox{}}%
	\def\@evenhead{\scriptsize\thepage \hfil \leftmark\mbox{}}%
	\def\@oddfoot{\parbox[t][\height][l]{\textwidth}{
			\vspace{-20pt}{\rule{\textwidth}{0.4pt}}\\ \footnotesize{\bf{\footnotesize\textcolor{red}{H. Naser, H. A. Hashim, and M. Ahmadi, "Quaternion-based Unscented Kalman Filter for Robust Wrench Estimation of Human-UAV Physical Interaction," Signal Processing, vol. 245, pp. 110582, 2026.}}} doi: \href{https://doi.org/10.1016/j.sigpro.2026.110582}{10.1016/j.sigpro.2026.110582}\\\\
			\noindent\makebox[\linewidth]
		}\hfil\hbox{}}%
	\def\@evenfoot{\MYfooter}}

\makeatother
\begin{document}
	\bstctlcite{IEEEexample:BSTcontrol}

\title{Quaternion-based Unscented Kalman Filter for Robust Wrench Estimation of Human-UAV Physical Interaction}

\author{Hussein Naser, Hashim A. Hashim, and Mojtaba Ahmadi% <-this % stops a space
	\thanks{This work was supported in part by the National Sciences and Engineering Research Council of Canada (NSERC), under the grants RGPIN-2022-04937.}
	\thanks{H. Naser, H. A. Hashim, and M. Ahmadi are with the Department of Mechanical
		and Aerospace Engineering, Carleton University, Ottawa, Ontario, K1S-5B6,
		Canada (e-mail: hhashim@carleton.ca).}
}

% \markboth{IEEE TRANSACTIONS ON INTELLIGENT TRANSPORTATION SYSTEMS, \today}{Hashim \MakeLowercase{\textit{et al.}}: Landmark and IMU Data Fusion: Systematic Convergence Geometric Nonlinear Observer for SLAM and Velocity Bias}

% \markboth{}{Hashim \MakeLowercase{\textit{et al.}}: Nonlinear Filter for Simultaneous Localization and Mapping on a Matrix Lie Group using IMU and Feature Measurements}

\maketitle
\begin{abstract}
This paper introduces an advanced Quaternion-based Unscented Kalman Filter (QUKF) for real-time, robust estimation of system states and external wrenches in assistive aerial payload transportation systems that engage in direct physical interaction. Unlike conventional filtering techniques, the proposed approach employs a unit-quaternion representation to inherently avoid singularities and ensure globally consistent, drift-free estimation of the platform's pose and interaction wrenches. A rigorous quaternion-based dynamic model is formulated to capture coupled translational and rotational dynamics under interaction forces. Building on this model, a comprehensive QUKF framework is established for state prediction, measurement updates, and external wrench estimation. The proposed formulation fully preserves the nonlinear characteristics of rotational motion, enabling more accurate and numerically stable estimation during physical interaction compared to linearized filtering schemes. Extensive simulations validate the effectiveness of the QUKF, showing significant improvements over the Extended Kalman Filter (EKF). Specifically, the QUKF achieved a 79.41\% reduction in Root Mean Squared Error (RMSE) for torque estimation, with average RMSE improvements of 79\% and 56\%, for position and angular rates, respectively. These findings demonstrate enhanced robustness to measurement noise and modeling uncertainties, providing a reliable foundation for safe, stable, and responsive human-UAV physical interaction in cooperative payload transportation tasks.
\end{abstract}


\section{Introduction}\label{sec1}

\subsection{Motivation}
\IEEEPARstart{A}{ssistive} Cooperative payload transportation using Unmanned Aerial Vehicles (UAVs) is a rapidly advancing field that utilizes multi-agent systems to carry and deliver payloads exceeding the capacity of a single vehicle \cite{Hash2025Avionics,wu2022collaborative}. This collaborative approach is vital in scenarios where traditional ground-based logistics are impractical, such as disaster-stricken regions, rugged terrains, or restricted urban environments \cite{chandran2024multi}. This approach enhances operational efficiency, reduces human risk, and provides access to otherwise inaccessible locations \cite{naser2025aerial}. The multi-agent systems can enable critical missions: for instance, delivering medical supplies in search and rescue operations \cite{cheema2022blockchain}, employing AI-based vision for wildfire sensing \cite{bouguettaya2022review}, and transporting heavy materials in elevated construction sites \cite{nwaogu2023application}.
However, the necessity of synchronized movement and stability demands advanced control algorithms and precise estimation of external interaction forces and torques \cite{naser2025aerial}. These factors are essential for preventing payload drops and ensuring safety and seamless operation, particularly when human operators are integrated into the control loop \cite{naser2025human}. Consequently, overcoming these coordination challenges in cooperative payload transportation marks a significant milestone in deploying UAV technology for complex, real-world applications.

\subsection{The Force Sensing Challenge in Aerial Systems \label{force_sensing_ghallenge}}

Conventionally, external wrenches are measured directly via 6D force-torque sensors \cite{berezny2025improving}. However, for aerial platforms, such hardware is often prohibitive \cite{tomic2014unified}. Beyond substantial economic costs that limit the scalability of multi-UAV systems, these sensors impose a severe mass penalty on weight-sensitive airframes, drastically reducing operational flight time \cite{Hash2025Avionics}. Furthermore, integration is technically demanding, requiring specialized mounting, signal conditioning, and power management, which increases system complexity and introduces additional failure modes \cite{barawkar2023force}. The need for frequent calibration to mitigate measurement drift further complicates their use. Consequently, estimation-based approaches that infer interaction wrenches from existing onboard sensors (e.g., IMUs and motor RPMs) are essential for developing practical, scalable human-UAV cooperative systems.

\subsection{Related Work in Sensorless Force Estimation}

Accurate estimation of interaction forces has emerged as a critical challenge in robotics, driven by the requirement for safe, robust, and adaptive control in collaborative environments \cite{park2024object}. Early research primarily focused on robotic manipulators, utilizing force-torque sensors for tasks such as assembly and human-robot collaboration \cite{mariotti2019admittance, li2024multi}. However, as detailed in Section \ref{force_sensing_ghallenge}, the integration of such hardware into aerial platforms is often precluded by weight, cost, and complexity constraints. These drawbacks have catalyzed a shift toward sensorless force estimation, where interaction wrenches are inferred from system dynamics, kinematics, or data-driven models.

Various observers and learning-based techniques have been explored to bypass physical sensing. Nonlinear observers have been proposed to estimate external disturbances, including human-applied forces, payload effects, and friction \cite{chen2000nonlinear, yuksel2014nonlinear, nikoobin2009lyapunov}. For instance, Rajappa et al. \cite{rajappa2017design} utilized residual-based estimators to decouple operator inputs from other disturbances. While computationally efficient, these observers often struggle with unmodeled dynamics and offer limited observability. Alternatively, learning-enhanced estimation approaches have also been explored to bypass physical sensing hardware \cite{berger2019deep}. For example, machine learning architectures, such as the adaptive recurrent neural network (RNN) developed by Sun et al. \cite{sun2024development}, offer high flexibility and are capable of capturing complex, unmodeled human-robot interaction dynamics. Recent study \cite{lu2025learning} have introduced learning-based approaches that utilize transformer architectures for interaction force estimation. However, despite their predictive power, the heavy training requirements, potential unpredictability outside of training distributions, and high computational overhead of deep learning models often render them impractical for time-critical applications like human-UAV cooperation.

Kalman-type filters provide a systematic framework by combining model predictions with sensor measurements \cite{cui2025variational, shao2024novel}; however, conventional linear Kalman filters are often restricted by their strict assumptions of linearity and Gaussian noise. To address these limitations, Extended Kalman Filters (EKFs) are frequently employed, yet they remain prone to linearization errors and Jacobian-related instabilities \cite{mckinnon2016unscented}. Building upon these nonlinear estimation techniques, the Unscented Kalman Filter (UKF) mitigates such issues by utilizing the Unscented Transformation (UT) to propagate sigma points through nonlinear dynamics, making it highly suitable for the complex nonlinearities inherent in aerial systems \cite{sorrentino2024ukf, ghanizadegan2024quaternion}. Despite these advantages, standard UKF applications, such as the square-root UKF used by Banks et al. \cite{banks2021physical} to estimate human-UAV interaction forces, often rely on Euler angles which introduce the risk of gimbal lock singularities \cite{hashim2021gps, hashim2023exponentially}. To ensure global nonsingularity, several geometrically consistent frameworks have been developed to preserve rotational constraints, including the Multiplicative EKF (MEKF) \cite{chang2023multiplicative}, Invariant EKF (IEKF) \cite{ko2018improvement}, and Lie Group-based UKFs on $SO(3)$ or $SE(3)$ \cite{sjoberg2021lie}. While these methods typically rely on local error-states and exponential mappings, our proposed QUKF directly propagates unit-quaternions within the unscented transform. This approach avoids singularities and preserves the full nonlinear kinematics of attitude evolution without the need for first-order linearization or complex group retractions, thereby simplifying the estimation of translational states, rotational states, and external interaction wrenches. From a theoretical perspective, the unscented transform captures higher-order moments of nonlinear quaternion dynamics more effectively than EKF-based schemes, leading to improved accuracy under strong coupling between rotational and translational dynamics \cite{ghanizadegan2024quaternion}. From a practical standpoint, this quaternion-based UKF offers reduced computational complexity and straightforward tuning, making it well-suited for real-time implementation on embedded UAV platforms during physical human-UAV interaction. This robustness is supported by recent work \cite{ghanizadegan2024quaternion} demonstrating that QUKFs outperform Euler-based formulations in GPS-denied navigation by maintaining a continuous attitude representation.

\subsection{Motivation and Contributions}

Collectively, previous research highlights the feasibility of sensorless force estimation and the efficacy of probabilistic filtering, particularly UKF and QUKF variants, in managing nonlinear, uncertain, and noisy dynamics. However, current literature focuses predominantly on ground manipulators or pure navigation tasks, with limited attention given to external wrench estimation for human-UAV cooperative payload transportation. While robust, quaternion-aware filters exist, their integration into a unified framework for real-time interaction force estimation in aerial systems remains an open challenge. Overcoming this gap is essential for implementing responsive admittance control and ensuring operational safety without the weight and cost penalties of physical sensors.

In this paper, we bridge this gap by proposing a novel Quaternion-based Unscented Kalman Filter (QUKF). This framework is designed to simultaneously estimate the full system state and external wrenches for an assistive aerial system during human-guided payload transportation. The key contributions of this work are as follows:

\begin{itemize}
	\item[C1.] Assistive Cooperative Framework: The design of a comprehensive framework for an assistive cooperative payload transportation that incorporates direct physical human guidance and interaction.
	\item[C2.] Quaternion-based Dynamic Modeling: The derivation of a rigorous quaternion-based dynamic model of the UAV-payload system, providing a robust mathematical foundation for both stability analysis and model-based estimation.
	\item[C3.]  Geometric QUKF Estimator: The development of a novel geometric QUKF for the concurrent estimation of navigation states and external interaction wrenches, effectively eliminating the requirement for 6-DoF force-torque sensors.
	\item[C4.] Comprehensive Validation: A high-fidelity simulation study integrating the QUKF estimator with a standard admittance controller to: (i) enable seamless physical human-UAV interaction, (ii) quantitatively validate estimation accuracy against baseline methods (e.g., EKF), and (iii) demonstrate a robust, singularity-free solution for all-attitude flight regimes.
\end{itemize}

The rest of the paper is organized as follows: Section \ref{preliminaries} elaborates on the notation used in this paper and provides mathematical preliminaries and orientation representation methods. Section \ref{Problem-Formulation} presents a comprehensive description of problem formulation and system modeling. Section \ref{kalman_filter} details the QUKF implementation. The simulation numerical results are provided in Section \ref{results-discussion}. Finally, Section \ref{Conclusion} concludes the article and outlines
potential future work.

\section{Preliminaries \label{preliminaries}}

\subsection{Notation}

The notation utilized throughout this paper is defined as follows: The sets of real and positive real numbers are denoted by $\mathbb{R}$ and $\mathbb{R}^{+}$, respectively, while $\mathbb{S}^{3}$ represents the three-unit sphere. The space of $n \times n$ real matrices is denoted by $\mathbb{R}^{n \times n}$. The terms $\mathbf{I}_{n}$ and $0_{n}$ represent the $n \times n$ identity and zero matrices. For a vector $p \in \mathbb{R}^{n}$, its transpose is denoted by $p^{\top}$ and its Euclidean norm is defined as $\|p\| = \sqrt{p^{\top}p}$. For any vector $p \in \mathbb{R}^{3}$, the corresponding skew-symmetric matrix $[p]_{\times} \in \mathfrak{so}(3)$ is defined as:
\begin{equation}
	\begin{bmatrix}p\end{bmatrix}_{\times}=\begin{bmatrix}0 & -p_{3} & p_{2}\\
		p_{3} & 0 & -p_{1}\\
		-p_{2} & p_{1} & 0
	\end{bmatrix}\in\mathfrak{so}(3),\hspace{1em}p=\begin{bmatrix}p_{1}\\
		p_{2}\\
		p_{3}
	\end{bmatrix}.\label{skew-symmetric_equ}
\end{equation}
The inverse skew-symmetric operator vex($\cdot$) is a mapping from
$\mathfrak{so}(3)$ in \eqref{skew-symmetric_equ} to $\mathbb{R}^{3}$,
($\text{vex}(\cdot):\mathfrak{so}(3)\rightarrow\mathbb{R}^{3}$),
such that \cite{ghanizadegan2024quaternion}:
\begin{equation}
	\text{vex}([p]_{\times})=p\in\mathbb{R}^{3}.\label{vex_equ}
\end{equation}
while the anti-symmetric projection operator $\mathcal{P}_{a}(\cdot)$
is a mapping from $\mathbb{R}^{3\times3}$ to $\mathfrak{so}(3)$,
($\mathcal{P}_{a}(\cdot):\mathbb{R}^{3\times3}\rightarrow\mathfrak{so}(3)$),
such that \cite{ghanizadegan2024quaternion}:
\begin{equation}
	\mathcal{P}_{a}(B)=\frac{1}{2}(B-B^{\top})\in\mathfrak{so}(3),~\forall B\in\mathbb{R}^{3\times3}.\label{anti_sym_proj_op_equ}
\end{equation}

\subsection{Representation of a rigid-body attitude in 3D space}

The attitude (orientation) of a rigid body moving in three-dimensional (3D) space can be described through several mathematical frameworks. These representations relate two distinct coordinate systems, a world inertial frame $\mathcal{I}_{f}=\{O,X,Y,Z\}$, which is global and Earth-fixed, and a body-fixed frame $\mathcal{B}_{f}=\{o,x_{b},y_{b},z_{b}\}$, attached to the Center-of-Mass (CoM) of the moving rigid-body. A brief description of some orientation representation methods will be given as follows:

\subsubsection{Special Orthogonal Group $SO(3)$}

The orientation is most fundamentally represented by a rotation matrix $R \in \mathbb{R}^{3 \times 3}$. This matrix belongs to the Special Orthogonal Group $SO(3)$, defined by:
\begin{equation}
	SO(3):=\{R\in\mathbb{R}^{3\times3}|det(R)=+1,~\text{and}~RR^{\top}=\mathbf{I}_{3}\},\label{rot_mat_equ}
\end{equation}
where $\det(\cdot)$ denotes the determinant. The rotation matrix $R$ provides a unique, universal representation for every physical orientation of the rigid body \cite{hashim2019special}.

\subsubsection{Euler angles}

Euler angles are widely adopted due to their intuitive visualization of rotations about the principal axes: roll ($\phi$) about $x$, pitch ($\theta$) about $y$, and yaw ($\psi$) about $z$ \cite{naser2025aerial}. However, despite their simplicity, they are mathematically limited by singularities known as gimbal lock, which can compromise stability in high-maneuverability applications \cite{hashim2019special}.

\subsubsection{Angle-axis}

In this method, the relative orientation between the inertial and body-fixed reference frames can also be expressed as a single rotation of angle $\alpha \in \mathbb{R}$ around a unit vector $b = [b_1, b_2, b_3]^{\top} \in \mathbb{S}^2$, where $\|b\|=\sqrt{b_{1}^{2}+b_{2}^{2}+b_{3}^{2}}=1$. While conceptually useful, this representation also encounters singularities in specific configurations. The parameters $(\alpha, b)$ can be extracted from a given rotation matrix $R$ via the mapping $(\alpha, b(R)): SO(3) \rightarrow \mathbb{S}^2 \times \mathbb{R}$) using the operators in \eqref{vex_equ} and \eqref{anti_sym_proj_op_equ} in the following relations \cite{hashim2019special}:
\begin{equation}
	\begin{aligned}b(R) & =\frac{1}{\sin(\alpha)}\text{vex}(\mathcal{P}_{a}(R))\in\mathbb{S}^{2},\\
		\alpha(R) & =\arccos\left(\frac{\text{Tr}(R)-1}{2}\right)\in\mathbb{R}.
	\end{aligned}
	\label{Rot_2_angle_axis_equ}
\end{equation}

\subsubsection{Rodriguez vector}

The Rodriguez vector $P = [p_1, p_2, p_3]^{\top} \in \mathbb{R}^3$ encodes rotation such that its direction aligns with the rotation axis and its magnitude is proportional to the rotation angle. It can be derived from the angle-axis form via the mapping ($P(\alpha,b):\mathbb{R}\times\mathbb{S}^{2}\rightarrow\mathbb{R}^{3}$) as follows:
\begin{equation}
	P(\alpha,b)=\alpha b\in\mathbb{R}^{3}.
	\label{Rodr_2_angle_axis_equ}
\end{equation}
Given Rodriguez vector, the corresponding rotation matrix $R(P)$ can be obtained via the mapping ($R(P)=\mathbb{R}^{3}\rightarrow SO(3)$)  as follows:
\begin{equation}
	R(P)=(\mathbf{I}_{3}+[P]_{\times})(\mathbf{I}_{3}-[P]_{\times})^{-1}\in SO(3).\label{Rodr_2_rot_equ}
\end{equation}
Like Euler angles and angle-axis representations, the Rodriguez vector representation is subject to mathematical singularities \cite{hashim2019special}.

\subsubsection{Quaternion \label{quat_subsection}}

Quaternions are four-dimensional hypercomplex numbers that provide a globally non-singular representation of 3D rotations. A quaternion $q=[q_{w},q_{v}^{\top}]^{\top}\in\mathbb{S}^{3}$ is defined as $q=q_{w}+q_{x}i+q_{y}j+q_{z}k$, where $i$, $j$, and $k$ refer to the standard basis vectors, $q_{w}\in\mathbb{R}$ is the scalar part, $q_{v}=[q_{x},q_{y},q_{z}]^{\top}\in\mathbb{R}^{3}$ is the vector part, and $q^{*}=q_{w}-q_{x}i-q_{y}j-q_{z}k$ is the complex conjugate of the quaternion $q$. A unit-quaternion is a quaternion satisfying $(\|q\|=1$ and $q^{-1}=q^{*})$, where $q^{-1}$ represents the inverse of $q$. Quaternions are generally preferred over Euler angles for UAV applications as they avoid gimbal lock and allow for smooth interpolation of attitudes. The quaternion multiplication and conjugation operations are fundamental for UAV orientation representation. The product of two quaternions $q_{1}$ and $q_{2}$, is given by:
\begin{equation}
	q_{1}\otimes q_{2}=\begin{bmatrix}q_{w_{1}}q_{w_{2}}-q_{v_{1}}^{\top}q_{v_{2}}\\
		q_{w_{1}}q_{v_{2}}+q_{w_{2}}q_{v_{1}}+[q_{v_{1}}]_{\times}q_{v_{2}}
	\end{bmatrix},\label{quat_multiply_equ}
\end{equation}
with $q\otimes q^{-1}=q_{I}$ and $q\otimes q_{I}=q$, where $q_{I}=[1,0,0,0]^{\top}$ is the quaternion identity. Quaternion multiplication is associative process but non-commutative. A quaternion is mapped to a rotation matrix $R(q) \in SO(3)$ via the mapping ($R(q):\mathbb{S}^{3}\rightarrow SO(3)$), as follows:
\begin{equation}
	\begin{aligned}R(q) & =(q_{w}^{2}-\|q_{v}\|^{2})\mathbf{I}_{3}+2q_{v}q_{v}^{\top}-2q_{w}[q_{v_{1}}]_{\times},\\
		& =\mathbf{I}_{3}+2q_{w}[q_{v_{1}}]_{\times}+2[q_{v_{1}}]_{\times}^{2},
	\end{aligned}
	\label{quat_2_rot_equ}
\end{equation}
Conversely, the rotation matrix can be remapped to a quaternion $q(R)$ via the mapping $(q(R):SO(3)\rightarrow\mathbb{S}^{3})$, as \cite{hashim2019special}:
\begin{equation}
	\begin{aligned}q(R) & =\begin{bmatrix}q_{w}\\
			q_{x}\\
			q_{y}\\
			q_{z}
		\end{bmatrix}=\begin{bmatrix}\frac{1}{2}\sqrt{1+R_{(1,1)}+R_{(2,2)}+R_{(3,3)}}\\
			\frac{1}{4q_{w}}(R_{(3,2)}-R_{(2,3)})\\
			\frac{1}{4q_{w}}(R_{(1,3)}-R_{(3,1)})\\
			\frac{1}{4q_{w}}(R_{(2,1)}-R_{(1,2)})
		\end{bmatrix}.\end{aligned}
	\label{rot_2_quat}
\end{equation}
Given a quaternion and/or a rotation vector $P$, and in view of equations \eqref{Rot_2_angle_axis_equ}, \eqref{Rodr_2_angle_axis_equ}, \eqref{Rodr_2_rot_equ}, \eqref{quat_2_rot_equ} and \eqref{rot_2_quat},
we obtain the following relationships:
\begin{equation}
	\begin{aligned}P(q)= & \alpha(R(q))*b(R(q))\in\mathbb{R}^{3},\\
		q(P)= & q(R(P))=\begin{bmatrix}\cos(\frac{\alpha}{2})\\
			b\sin(\frac{\alpha}{2})
		\end{bmatrix}\in\mathbb{S}^{3}.
	\end{aligned}
	\label{Vec_2_quat_2_vec_equ}
\end{equation}

\subsubsection{Summation, subtraction, and weighted average}

Direct summation and subtraction are not applicable between quaternions
and rotation vectors. In view of equations \eqref{quat_multiply_equ} and \eqref{Vec_2_quat_2_vec_equ}, one has \cite{ghanizadegan2024quaternion}:
\begin{equation}
	\begin{aligned}q\oplus P & :=q(P)\otimes q\in\mathbb{S}^{3},\\
		q\ominus P & :=q(P)^{-1}\otimes q\in\mathbb{S}^{3},
	\end{aligned}
	\label{sum_sub_equ}
\end{equation}
where $\oplus$ and $\ominus$ represent the quaternion $(q\in\mathbb{S}^{3})$ and vector $(P\in\mathbb{R}^{3})$ side-dependent summation and subtraction operators, respectively. Using equation \eqref{quat_multiply_equ},
\eqref{Vec_2_quat_2_vec_equ} and \eqref{sum_sub_equ}, one can subtract two subsequent quaternions and map the result to a rotation vector at the same time ($\mathbb{S}^{3}\times\mathbb{S}^{3}\rightarrow\mathbb{R}^{3}$), as follows:
\begin{equation}
	\begin{aligned}q_{1}\ominus q_{2} & :=P(q_{1}\otimes q_{2}^{-1})\in\mathbb{R}^{3}.\end{aligned}
	\label{Quat_sub_equ}
\end{equation}
The resultant rotation vector $P(q_{1}\otimes q_{2}^{-1})$ in \eqref{Quat_sub_equ}
physically means the orientation error between the two quaternions.
The weighted average of a set of scalar weights $S_{W}=\{\text{w}_{i}\}$
and quaternions $S_{q}=\{q_{i}\}$ is given as follows \cite{hashim2018nonlinear,ghanizadegan2024quaternion}:
\begin{equation}
	\operatorname{WAve}(S_{q},S_{W})=\text{Eigvec}(\mathbb{A})\in\mathbb{S}^{3},\label{weighted_avrg_equ}
\end{equation}
where $\mathbb{A}$ is a matrix given as $\mathbb{A}(S_{q},S_{W})=\sum\text{w}_{i}q_{i}q_{i}^{\top}\in\mathbb{R}^{4\times4}$
and $\text{Eigvec}(\mathbb{A})$ is the unit eigenvector of the matrix
$\mathbb{A}$ corresponding to its maximum eigenvalue. 

\begin{table}[h!]
	\centering{}\caption{\label{tab:general_symbols} Nomenclature.}
	\begin{tabular}{ll}
		\hline 
		Symbol  & Definition\tabularnewline
		\hline 
		\hline 
		$\mathcal{I}_{f}$,$L_{f}$,$\mathcal{B}_{fi}$  & Inertial- frame, payload body-frame, $i^{th}$ UAV body-frame.\tabularnewline
		$r_{i},v_{i},\dot{v}_{i}$  & $i^{th}$ UAV position, velocity, and acceleration in $\mathcal{I}_{f}$.\tabularnewline
		$r_{l}$, $v_{l}$, $\dot{v}_{l}$  & Payload position, velocity, acceleration in $\mathcal{I}_{f}$. \tabularnewline
		$r$, $v$, $\dot{v}$  & System position, velocity, and acceleration in $\mathcal{I}_{f}$. \tabularnewline
		$m_{i},\mathcal{J}_{i}$  & $i^{th}$ UAV mass and moment of inertia.\tabularnewline
		$m_{l}$, $\mathcal{J}_{l}$  & Payload mass and moment of inertia.\tabularnewline
		$m_{s}$, $\mathcal{J}$  & Total mass and moment of inertia of entire system. \tabularnewline
		$g$  & Gravity.\tabularnewline
		$\varpi$  & Angular speed of each quadrotor's rotor.\tabularnewline
		$\iota$  & Distance from quadrotor's CoM to the rotor's axis.\tabularnewline
		$k_{t},k_{m}$  & Thrust and drag constants.\tabularnewline
		$\mathfrak{F}_{i},\mathfrak{T}_{i}$  & $i^{th}$ UAV force and torque exerted on payload.\tabularnewline
		$SO(3)$  & Special Orthogonal Group.\tabularnewline
		$\mathfrak{so}(3)$  & Space of $(3\times3)$ skew-symmetric matrices.\tabularnewline
		$q$  & Quaternion $q\in\mathbb{S}^{3}$.\tabularnewline
		$R(q)$  & Rotation matrix $R(q)\in SO(3)$.\tabularnewline
		$\omega$, $\dot{\omega}$  & System angular velocity and acceleration in body-frame.\tabularnewline
		$\mathcal{F}_{th}$  & Total thrust produced by two quadrotors. \tabularnewline
		$\mathcal{U}_{\tau}$  & Total (rolling, pitching, yawing) moments of the system. \tabularnewline
		$\tau_{h},\hat{\tau}_{h}$  & External physical interaction wrench and its estimate.\tabularnewline
		$x_{k},\hat{x}_{k}$  & State vector and its estimate at time $k$.\tabularnewline
		$N$  & State vector length.\tabularnewline
		$\text{w}_{k},\text{v}_{k}$  & Process and measurement noise vectors at time $k$.\tabularnewline
		$\mathsf{Q},\mathsf{R}$  & Process and measurement covariance matrices.\tabularnewline
		$u_{k}$  & Control input vector at time $k$.\tabularnewline
		$y_{k}$  & Output vector at time $k$.\tabularnewline
		$\mathcal{L}_{k}$  & Sigma points matrix at time $k$.\tabularnewline
		$K$  & Kalman gain matrix at time $k$.\tabularnewline
		$P$  & Error covariance matrix.\tabularnewline
		\hline 
	\end{tabular}
\end{table}

\section{Problem formulation \label{Problem-Formulation}}

In human-UAV cooperative payload transportation, accurate estimation of external wrenches is critical for operational safety. This task is inherently challenging due to the highly nonlinear dynamics of the aerial platforms and the stochastic nature of human interaction. To address these challenges, this work proposes an innovative, sensorless method for estimating interaction forces and torques. The system under study comprises two quadrotors collaboratively transporting a rigidly attached payload.

Through this setup, a human operator intuitively guides the formation by applying physical forces directly to the payload. These interaction wrenches are estimated online using a Quaternion-based Unscented Kalman Filter (QUKF), which avoids the weight and complexity of physical force-torque sensors. The resulting estimates are processed by an admittance controller to translate human physical guidance into precise 3D motion commands. For the purpose of control design, the multi-UAV-payload assembly is modeled as a single integrated rigid body. The notation and symbols employed throughout this manuscript are detailed in Table \ref{tab:general_symbols}.

\subsection{System Model}

For the system modeling, we define a global inertial frame,
$\mathcal{I}_{f}=\{O,X,Y,Z\}$, with a positive $Z$-axis upward.
A body-fixed frame $L_{f}=\{o,x_{l},y_{l},z_{l}\}$ is attached to
the payload's center of mass (CoM), and a frame $\mathcal{B}_{fi}=\{o_{i},x_{b_{i}},y_{b_{i}},z_{b_{i}}\}$
is attached to the CoM of each quadrotor $(i=1,2)$; all the body-fixed frames
share the same upward $z$-axis orientation. The position and velocity
of the $i^{th}$ quadrotor in $\mathcal{I}_{f}$ are given by $r_{i}=[x_{i},y_{i},z_{i}]^{\top}$
and $v_{i}=\dot{r}_{i}$, respectively. Due to the rigid connection,
the system's rotational dynamics are described by a single unit-quaternion
$q=[q_{w},q_{v}^{\top}]^{\top}\in\mathbb{S}^{3}$ and a common angular
velocity vector $\omega=[p,q,r]^{\top}$ expressed in the body frame.

\begin{figure}[tb!]
	\centering{}\includegraphics[clip,width=0.85\columnwidth]{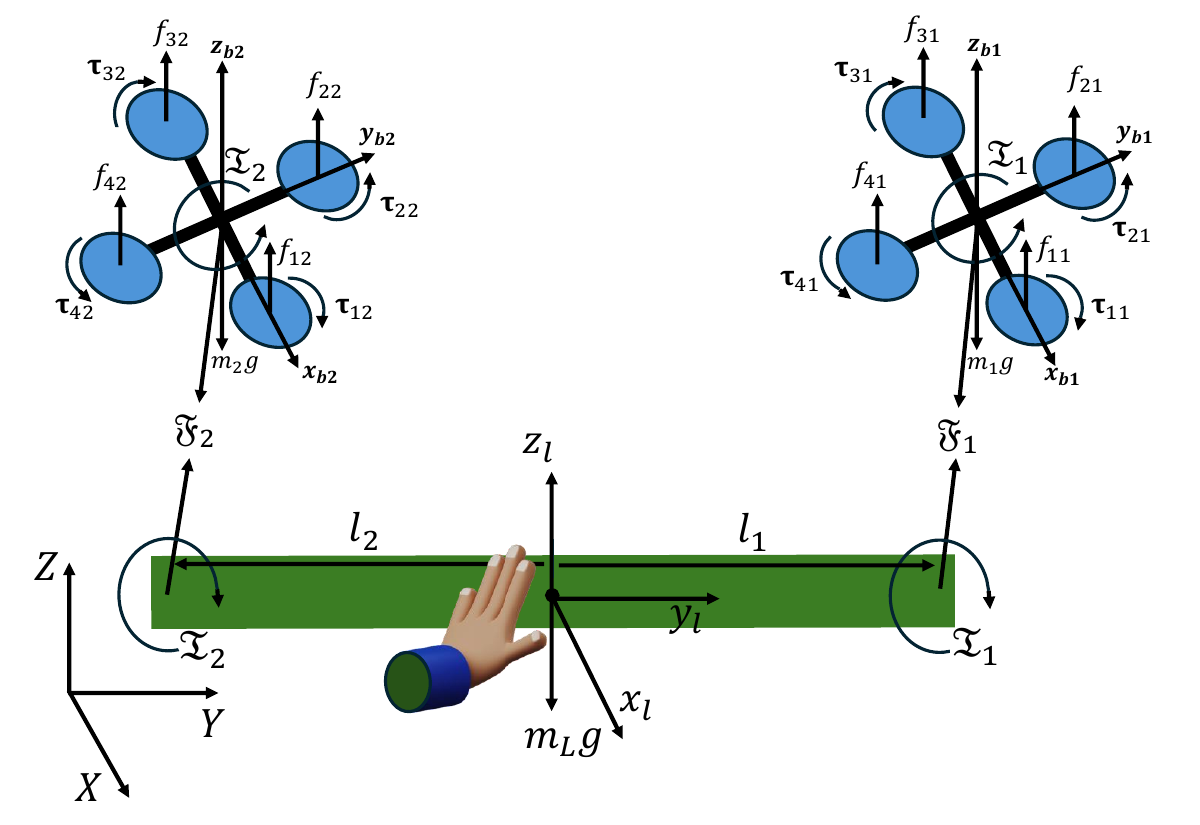}
	\caption{Free body diagram of the system components.}
	\label{Fig1:system_FBD} 
\end{figure}

\subsubsection{Quadrotor Dynamics}

Referring to Figure \ref{Fig1:system_FBD}, the translational and angular dynamics of the $i^{th}$ quadrotor are modeled using Newton-Euler formulation, as follows \cite{hashim2023exponentially,hashim2023observer}:
\begin{equation}
	\begin{aligned}\dot{q} & =\frac{1}{2}\Xi(\omega)q,\\
		\dot{r}_{i} & =v_{i},\\
		\dot{v}_{i} & =R(q)\frac{u_{i_{1}}}{m_{i}}E_{z}-gE_{z}-\frac{\mathfrak{F}_{i}}{m_{i}},\\
		\dot{\omega} & =\mathcal{J}_{i}^{-1}(u_{i\tau}-\omega\times\mathcal{J}_{i}\omega-\mathfrak{T}_{i}),
	\end{aligned}
	\label{trans_rot_quad_equ}
\end{equation}
where
\[
\Xi(\omega)=\begin{bmatrix}0 & -\omega^{\top}\\
	\omega & -\begin{bmatrix}\omega\end{bmatrix}_{\times}
\end{bmatrix}\in\mathbb{R}^{4\times4},
\]
$m_{i}\in\mathbb{R}$ and $\mathcal{J}_{i}\in\mathbb{R}^{3\times3}$
are the mass and inertia matrix; $\dot{v}_{i},\dot{\omega}\in\mathbb{R}^{3}$
are the translational and angular accelerations; $u_{i_{1}}\in\mathbb{R}$
is the generated thrust measured in the body-fixed reference frame;
$g\in\mathbb{R}$ is gravity; $E_{z}=[0,0,1]^{\top}$ is a unit vector
along $z$-axis; $u_{i\tau}=[u_{i_{2}},u_{i_{3}},u_{i_{4}}]^{\top}\in\mathbb{R}^{3}$
are the roll, pitch, and yaw moments about $x_{b_{i}},y_{b_{i}},z_{b_{i}}$;
and $\mathfrak{F}_{i},\mathfrak{T}_{i}\in\mathbb{R}^{3}$ are the
interaction force and torque due to the payload connection. The generated thrust and moments of the $i^{th}$ quadrotor are given
as follows:
{\small
\begin{equation}
	\begin{bmatrix}u_{i_{1}}\\
		u_{i\tau}
	\end{bmatrix} =\begin{bmatrix}f_{i_{1}}+f_{i_{2}}+f_{i_{3}}+f_{i_{4}}\\
		\iota(f_{i_{2}}-f_{i_{4}})\\
		\iota(f_{i_{3}}-f_{i_{1}})\\
		\tau_{i_{1}}-\tau_{i_{2}}+\tau_{i_{3}}-\tau_{i_{4}}
	\end{bmatrix}
	=\begin{bmatrix}1 & 1 & 1 & 1\\
		0 & \iota & 0 & -\iota\\
		-\iota & 0 & \iota & 0\\
		\nu & -\nu & \nu & -\nu
	\end{bmatrix}\begin{bmatrix}f_{i_{1}}\\
		f_{i_{2}}\\
		f_{i_{3}}\\
		f_{i_{4}}
	\end{bmatrix}.
	\label{total_i_thrust_moments_equ}
\end{equation} }
Each rotor of the $i^{th}$ quadrotor generates a thrust $(f_{ij}=k_{t}\varpi_{ij}^{2})$,
which is proportional to the squared rotor's angular speed $\varpi_{ij}$,
where $k_{t}$ is the thrust constant. Similarly, the moment produced
by each rotor is $(\tau_{ij}=k_{m}\varpi_{ij}^{2})$, with $k_{m}$
being the drag constant, and $\nu=k_{m}/k_{t}$. The parameter $\iota$
represents the distance from the quadrotor's CoM to the axis of rotation
of each rotor. The parameter $\iota$ is assumed to be the same for
all rotors.

\subsubsection{Payload Dynamics}

Referring to Figure \ref{Fig1:system_FBD}, let the payload CoM position
and velocity in $\mathcal{I}_{f}$ be $r_{l}=[x_{l},y_{l},z_{l}]^{\top}$
and $v_{l}=\dot{r}_{l}$, respectively. The payload translational
and rotational dynamics are then given by:
\begin{equation}
	\begin{aligned}\dot{q} & =\frac{1}{2}\Xi(\omega)q,\\
		\dot{r}_{l} & =v_{l},\\
		\dot{v}_{l} & =\frac{(\mathfrak{F}_{1}+\mathfrak{F}_{2})}{m_{l}}-gE_{z},\\
		\dot{\omega} & =\mathcal{J}_{l}^{-1}\Big((\mathfrak{T}_{1}+\mathfrak{T}_{2})-\omega\times\mathcal{J}_{l}\omega+(l_{1}\times\mathfrak{F}_{1})+(l_{2}\times\mathfrak{F}_{2})\Big),
	\end{aligned}
	\label{trans_rot_payload_equ}
\end{equation}
where $m_{l}\in\mathbb{R}$ denotes the payload mass,
and $\mathcal{J}_{l}\in\mathbb{R}^{3\times3}$ is its inertia matrix.
The translational acceleration is $\dot{v}_{l}\in\mathbb{R}^{3}$.
The forces $(\mathfrak{F}_{1},\mathfrak{F}_{2})\in\mathbb{R}^{3}$
and torques $(\mathfrak{T}_{1},\mathfrak{T}_{2})\in\mathbb{R}^{3}$
are applied by the quadrotors at the rigid connection points. The
corresponding force moments, $(l_{1}\times\mathfrak{F}_{1})$ and
$(l_{2}\times\mathfrak{F}_{2})$, depend on the vector $(l_{i}=r_{i}-r_{l})$,
where $r_{i}$ and $r_{l}$ are the CoM position vector of the $i^{th}$
quadrotor and the payload, respectively.

\subsubsection{Dynamics of the entire Aerial system}

To derive the dynamic equations of the entire system, we adopt the
following assumptions.

\begin{assum}\label{as:1} \textcolor{white}{.} 
	\begin{itemize}
		\item[A.] The system consists of three rigid components connected rigidly,
		each with distinct physical properties. 
		\item[B.] The two quadrotors are identical and share the same geometric and
		physical characteristics. 
		\item[C.] The entire system is symmetric about the $X$ and $Y$ axes. 
		\item[D.] The aerodynamic drag forces and torques are neglected. 
		\item[E.] The payload is modeled as a circular beam with mass $m_{l}$, length
		$\mathfrak{L}$, and cross-sectional radius $a_{l}$. 
	\end{itemize}
\end{assum}

Let $r=[x,y,z]^{\top}$ and $v=\dot{r}$ denote the position and velocity
of the system CoM in $\mathcal{I}_{f}$. Based on Assumption \ref{as:1}
and referring to Figure \ref{Fig1:system_FBD}, the system dynamic
model is derived by combining the dynamics of the quadrotors in \eqref{trans_rot_quad_equ}
and the payload in \eqref{trans_rot_payload_equ}, as follows:
\begin{equation}
	\begin{aligned}\dot{q} & =\frac{1}{2}\Xi(\omega)q,\\
		\dot{r} & =v,\\
		\dot{v} & =R(q)\frac{\mathcal{F}_{th}}{m_{s}}E_{z}-gE_{z}+\frac{\mathcal{F}_{h}}{m_{s}},\\
		\dot{\omega} & =\mathcal{J}^{-1}(\mathcal{U}_{\tau}-\omega\times\mathcal{J}\omega+M_{h}),
	\end{aligned}
	\label{trans_rot_sys_equ}
\end{equation}
where $(m_{s}=\sum_{i=1}^{2}m_{i}+m_{l})\in\mathbb{R}$
is the total mass of the system, and $\mathcal{J}\in\mathbb{R}^{3\times3}$
its total inertia. The translational acceleration is $\dot{v}\in\mathbb{R}^{3}$.
External human interaction is modeled by force $\mathcal{F}_{h}\in\mathbb{R}^{3}$
and torque $M_{h}\in\mathbb{R}^{3}$. The quadrotors generate a total
thrust $\mathcal{F}_{th}\in\mathbb{R}$ and control moments $\mathcal{U}_{\tau}\in\mathbb{R}^{3}$,
expressed in the system body-fixed frame. Since each quadrotor produces
inputs in its own local frame, their contributions must be mapped
to the system dynamics as follows:
\begin{equation}
	\begin{bmatrix}\mathcal{F}_{th}\\
		\mathcal{U}_{\tau}
	\end{bmatrix}=\mathbb{C}u_{c},\label{contro_input_equ}
\end{equation}
where $u_{c}\in\mathbb{R}^{4N}$ represents the quadrotors' control
inputs vector.
\begin{equation}
	u_{c}=[u_{11},u_{12},u_{13},u_{14},u_{21},u_{22},u_{23},u_{24}]^{\top},\label{u_inputs}
\end{equation}
and $\mathbb{C}\in\mathbb{R}^{4\times4N}$ represents the configuration
matrix that is constant and depends on the system configuration. For
the rigidly connected system in this work, $\mathbb{C}$ is given
as follows:
\begin{equation}
	\mathbb{C}=\begin{bmatrix}1 & 0 & 0 & 0 & 1 & 0 & 0 & 0\\
		l_{1}(2) & 1 & 0 & 0 & l_{2}(2) & 1 & 0 & 0\\
		-l_{1}(1) & 0 & 1 & 0 & -l_{2}(1) & 0 & 1 & 0\\
		0 & 0 & 0 & 1 & 0 & 0 & 0 & 1
	\end{bmatrix}.\label{b_matrix_equ}
\end{equation}

\begin{rem}
	\label{rem_1}In the proposed control scheme, $[\mathcal{F}_{th},\mathcal{U}_{\tau}]^{\top}$
	is first calculated and then utilized to determine $(u_{c})$ in \eqref{u_inputs},
	allowing the quadrotors to achieve the desired tracking control. Notably,
	the system in \eqref{contro_input_equ} is underdetermined, as it
	provides only four equations for eight unknown variables. To resolve
	this, the solutions are optimized by minimizing a cost function $\mathfrak{J}(u_{c}):\mathbb{R}^{8}\rightarrow\mathbb{R}$,
	defined as follows:
	\begin{equation}
		u_{c}^{*}=\text{argmin}\{\mathfrak{J}|[\mathcal{F}_{th},\mathcal{U}_{\tau}]^{\top}=\mathbb{C}u_{c}\},\label{optim_fun_equ}
	\end{equation}
\end{rem}
here, the cost function $\mathfrak{J}$ in \eqref{optim_fun_equ}
is defined as follows:
\begin{equation}
	\mathfrak{J}=\sum_{i=1}^{2}\sum_{j=1}^{4}\kappa_{ij}u_{ij}^{2},\label{cost_fun_equ}
\end{equation}
where $\kappa_{ij}$ represents the coefficients of the cost function,
which can be used to construct a matrix $\Lambda\in\mathbb{R}^{8\times8}$.
Consequently, the cost function $\mathfrak{J}$ in \eqref{cost_fun_equ}
can be reformulated as $(\mathfrak{J}=\|\Lambda u_{c}\|_{2}^{2})$,
where $\Lambda$ is defined as follows:
\begin{equation}
	\Lambda=\sqrt{diag(\kappa_{ij})},\hspace{1em}i=1,2,~~j=1,...,4.\label{h_matrix_equ}
\end{equation}
Utilizing $\mathbb{C}$ from \eqref{b_matrix_equ} and $\Lambda$
from \eqref{h_matrix_equ}, an optimal solution can be computed using
the Moore-Penrose inverse as follows \cite{mellinger2013cooperative}:
\begin{equation}
	\begin{split}u_{c}^{*} & =\Lambda^{-1}(\mathbb{C}\Lambda^{-1})^{+}[\mathcal{F}_{th},\mathcal{U}_{\tau}]^{\top},\\
		& =\Lambda^{-2}\mathbb{C}^{\top}(\mathbb{C}\Lambda^{-2}\mathbb{C}^{\top})^{-1}[\mathcal{F}_{th},\mathcal{U}_{\tau}]^{\top},
	\end{split}
	\label{optim_cont_final_equ}
\end{equation}
here, the superscript $+$ denotes the Moore-Penrose inverse, and
the optimized solution $u_{c}^{*}$ in \eqref{optim_cont_final_equ}
satisfies the conditions outlined in Remark \ref{rem_1}.

\subsection{Assistive Control System Architecture\label{control_section}}

\begin{figure}
	\centering{}\includegraphics[width=0.99\columnwidth]{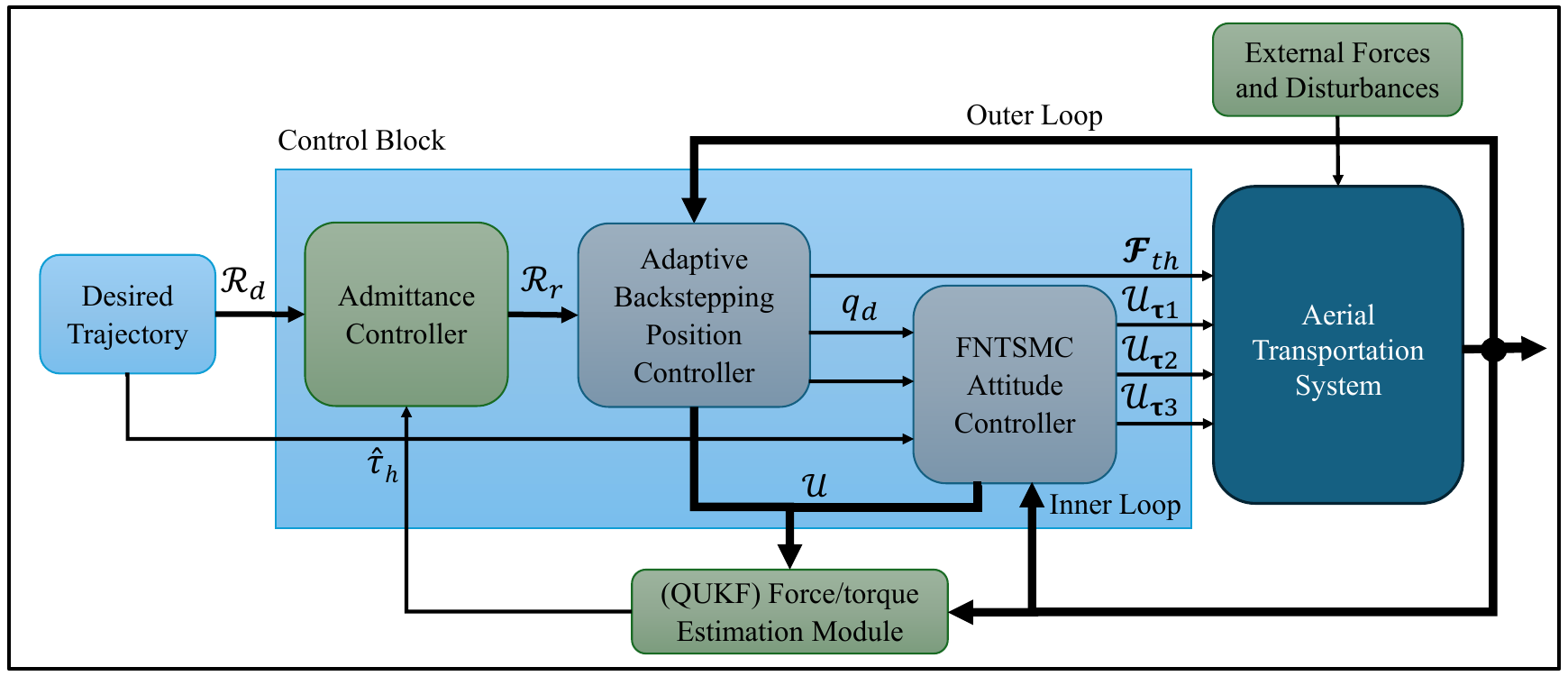}
	\caption{Schematic block diagram of the control system.}
	\label{Fig2:control_SBD} 
\end{figure}

To facilitate intuitive physical human-UAV interaction, the control architecture integrates three primary objectives: (i) human-led guidance, (ii) reference trajectory tracking, and (iii) stabilization of the cooperative payload transportation system. The position subsystem is stabilized via a Backstepping control law, while a Fast Nonsingular Terminal Sliding Mode Controller (FNTSMC) is employed to ensure robust, fast attitude regulation. Furthermore, an admittance control layer is incorporated to modulate the interaction forces, enabling seamless and intuitive cooperation between the human operator and the UAV. The integrated control framework is depicted in Figure \ref{Fig2:control_SBD}. For a comprehensive treatment of the controller's design and formal stability derivations, the reader is referred to \cite{naser2025human, naser2025aerial}.

\subsection{Estimation of external human-exerted wrench}

For estimation purposes and in order to find a unified expression
for the interaction wrench, the dynamic model in \eqref{trans_rot_sys_equ}
can be written in one compact form as follows:
\begin{equation}
	\tau_{h}=M\ddot{\chi}+G(\dot{\chi})+W(\chi)u,\label{estimation_wrench_equ}
\end{equation}
where $\tau_{h}=[\mathcal{F}_{h}^{\top},M_{h}^{\top}]^{\top}\in\mathbb{R}^{6}$
represents the external interaction wrench, $\ddot{\chi}=[\dot{v}^{\top},\dot{\omega}^{\top}]^{\top}\in\mathbb{R}^{6}$
is the generalized acceleration vector, $u=[\mathcal{F}_{th},\mathcal{U}_{\tau}^{\top}]^{\top}\in\mathbb{R}^{4}$
is the vector of control inputs, $M\in\mathbb{R}^{6\times6}$ is the
positive definite inertia matrix, $G(\dot{\chi})\in\mathbb{R}^{6}$
is a vector contains the gravity and the cross product terms in \eqref{trans_rot_sys_equ},
$W(\chi)\in\mathbb{R}^{6\times4}$ is the control input coefficients
matrix. These matrices are given as follows:
\begin{align*}
	M= & \begin{bmatrix}m_{s}\mathbf{I}_{3} & 0_{3}\\
		0_{3} & \mathcal{J}
	\end{bmatrix}\in\mathbb{R}^{6\times6},\hspace{1em}G=\begin{bmatrix}m_{s}gE_{z}\\
		\omega\times\mathcal{J}\omega
	\end{bmatrix}\in\mathbb{R}^{6},\\
	W(\chi)= & \begin{bmatrix}-R(q)E_{z} & 0_{3}\\
		0_{3\times1} & -\mathbf{I}_{3}
	\end{bmatrix}\in\mathbb{R}^{6\times4},
\end{align*}
where $\mathbf{I}_{3}$ and $0_{3}$ are identity and zero matrices
with dimension of ($3\times3$), respectively. Let $\hat{\tau}_{h}=[\hat{\mathcal{F}}_{h}^{\top},\hat{M}_{h}^{\top}]^{\top}\in\mathbb{R}^{6}$
represents the estimation of the external interaction wrench, and
let its rate of change over time be given as:
\begin{equation}
	\dot{\hat{\tau}}_{h}=A\big(\tau_{h}-\hat{\tau}_{h}\big)=-A\hat{\tau}_{h}+A\Big(M\ddot{\chi}+G(\dot{\chi})+W(\chi)u\Big),\label{wrench_observer_equ}
\end{equation}
where $A\in\mathbb{R}^{6\times6}$ is a matrix that will be designed
to guarantee that $(\hat{\tau}_{h}\rightarrow\tau_{h})$. Since we
have no prior knowledge about the external interaction wrench and
its derivative except that it will not be rapidly varying during the
physical interaction between the assistive aerial system and the human
operator, it is convenient to assume the following:
\begin{equation}
	\dot{\tau}_{h}=0.\label{derivative_zero_equ}
\end{equation}
Let us define the wrench error as follows:
\begin{equation}
	e=\tau_{h}-\hat{\tau}_{h},\label{error_observer_equ}
\end{equation}
Differentiating \eqref{error_observer_equ} with respect to time and
use \eqref{derivative_zero_equ}, results in:
\begin{equation}
	\dot{e}=\dot{\tau}_{h}-\dot{\hat{\tau}}_{h}=A\hat{\tau}_{h}-A\tau_{h}.\label{error_dot_observer_equ}
\end{equation}
Based on \eqref{error_observer_equ}, equation \eqref{error_dot_observer_equ}
can be expressed in the form of error dynamics as follows:
\begin{equation}
	\dot{e}+Ae=0,\label{error_dot_dynamic_observer_equ}
\end{equation}
From \eqref{error_dot_dynamic_observer_equ}, the
error convergence $(e\to0)$ can be guaranteed by an appropriate design
of $A$. As seen in \eqref{wrench_observer_equ}, computing $\dot{\hat{\tau}}_{h}$
requires $(\chi,\dot{\chi},\ddot{\chi})$. While $\chi$ and $\dot{\chi}$
can be measured or estimated using standard methods, $\ddot{\chi}$
(particularly the angular acceleration) is typically not available.
Therefore, an acceleration-free formulation of \eqref{wrench_observer_equ}
is required. To this end, the following auxiliary vector is introduced
\cite{zhong2024prototype}:
\begin{equation}
	\Upsilon=\hat{\tau}_{h}-\Gamma(\dot{\chi}),\label{auxiliary_equ}
\end{equation}
Differentiating \eqref{auxiliary_equ} with respect to time results
in:
\begin{equation}
	\dot{\hat{\tau}}_{h}=\dot{\Upsilon}+\frac{\partial\Gamma(\dot{\chi})}{\partial\dot{\chi}}\ddot{\chi}.\label{auxiliary_dot_equ}
\end{equation}
Using \eqref{auxiliary_equ} and \eqref{auxiliary_dot_equ} in equation
\eqref{wrench_observer_equ}, results in:
\begin{equation}
	\dot{\Upsilon}+\frac{\partial\Gamma(\dot{\chi})}{\partial\dot{\chi}}\ddot{\chi}=-A\big(\Upsilon+\Gamma(\dot{\chi})\big)+A\big(M\ddot{\chi}+G(\dot{\chi})+W(\chi)u\big),\label{auxiliary_wrench_observer_equ}
\end{equation}
Simplifying \eqref{auxiliary_wrench_observer_equ}, the acceleration-free
dynamics of the external interaction wrench is expressed as follows:
\begin{equation}
	\begin{aligned}\dot{\Upsilon} & =-A\Upsilon+A\Big(G(\dot{\chi})+W(\chi)u-\Gamma(\dot{\chi})\Big),\\
		\hat{\tau}_{h} & =\Upsilon+\Gamma(\dot{\chi}),
	\end{aligned}
	\label{auxiliary_wrench_observer_dynamics_equ}
\end{equation}
where $(\partial\Gamma(\dot{\chi})/\partial\dot{\chi})$ is chosen
as follows:
\begin{equation}
	\frac{\partial\Gamma(\dot{\chi})}{\partial\dot{\chi}}=AM.\label{chosen_auxiliary_equ}
\end{equation}
From \eqref{error_dot_dynamic_observer_equ} and
\eqref{chosen_auxiliary_equ}, the matrix $A$ must be designed to
guarantee the asymptotic stability of the error dynamics. We proceed
by considering the following:
\[
\Gamma(\dot{\chi})=\delta\dot{\chi}\hspace{1em}\Rightarrow\hspace{1em}A=\delta M^{-1},
\]
where $\delta\in\mathbb{R}^{+}$ is a positive tunable gain. While a comprehensive treatment of the overall control system's stability can be found in \cite{naser2025aerial, naser2025human, naser2026agno}, the stability of the proposed external wrench estimation scheme is governed by the error dynamics derived in equation \eqref{error_dot_dynamic_observer_equ}. The stability margin of this estimator is intrinsically tied to the design matrix $A$. Since $\delta$ is a positive gain and $M$ is the positive definite inertia matrix, the error dynamics are guaranteed to be asymptotically stable. This ensures that the estimated wrench ($\hat{\tau}_{h}$) consistently converges to the actual external interaction wrench ($\tau_{h}$) without sustained oscillations during physical interaction.

\section{Quaternion-based Unscented Kalman Filter (QUKF) \label{kalman_filter}}

The UKF is an advanced estimation algorithm designed for nonlinear systems with stochastic noise. Unlike the traditional Extended Kalman Filter (EKF), which relies on first-order linearization, the UKF addresses nonlinearities more accurately through deterministic sampling. In this work, the QUKF is designed to estimate the full augmented state vector, comprising position, orientation, linear/angular velocities, and the external interaction wrench, given noisy position, orientation (quaternion) and angular velocity data. In stochastic estimation problems, it is often assumed that the process and measurement noise are added linearly as follows:
\begin{equation}
	\begin{aligned}x_{k} & =\operatorname{f}(x_{k-1},u_{k-1})+\text{w}_{k-1},\\
		y_{k} & =\operatorname{h}(x_{k})+\text{v}_{k},
	\end{aligned}
	\label{nonlinear_tran_fun_equ}
\end{equation}
where $x_{k}$, $y_{k}$, and $u_{k}$ represent the state, output, and input vectors at discrete time $k$. The functions $\operatorname{f}$ and $\operatorname{h}$ denote the nonlinear state transition and observation mappings, respectively. The noise terms $\text{w}_{k}$ and $\text{v}_{k}$ are assumed to be independent, zero-mean, white Gaussian processes with covariance matrices $\mathsf{Q}$ and $\mathsf{R}$ \cite{rhudy2013understanding}:
\begin{equation}
	\begin{aligned} & \text{w}_{k}\sim\mathcal{N}(0,\mathsf{Q}_{k}),\\
		& \text{v}_{k}\sim\mathcal{N}(0,\mathsf{R}_{k}),\\
		& E[\text{w}_{k}\text{v}_{k}^{\top}]=0.
	\end{aligned}
	\label{process_measure_noise_mat_eaq}
\end{equation}
In this formulation, the process noise $\text{w}_{k}$ accounts for unmodeled dynamics, aerodynamic uncertainties, and variations in human-applied forces, while the measurement noise $\text{v}_{k}$ captures sensor-level imperfections. By explicitly incorporating these sources of noise into the UKF prediction and update steps, their effects are propagated through the nonlinear dynamics via the unscented transformation. In view of \eqref{nonlinear_tran_fun_equ}, the state transition function that models the relationship between the current state, the previous state, and the input vectors can be derived by augmenting the nonlinear system dynamics in \eqref{trans_rot_sys_equ} with the external wrench evolution in \eqref{auxiliary_wrench_observer_dynamics_equ}. This results in the following continuous-time augmented system:
\begin{equation}
	\dot{x}=\operatorname{f}(x,u),\label{equ:augmented_dynamics}
\end{equation}
with the augmented state vector defined as $x=[q,r,v,\omega,\Upsilon]^{\top}\in\mathbb{R}^{19}$, where $q\in\mathbb{S}^{3}$, $r,v,\omega\in\mathbb{R}^{3}$, and $\Upsilon\in\mathbb{R}^{6}$. Unlike conventional methods that treat external disturbances as lumped noise, this framework embeds the external wrench dynamics into the state vector to capture the inherent coupling between the platform's attitude and external interactions during the sigma-point propagation step. The right-hand side of \eqref{equ:augmented_dynamics} is expressed as:
\begin{equation}
	\operatorname{f}(x,u)=\begin{bmatrix}\dot{q}\\
		\dot{r}\\
		\dot{v}\\
		\dot{\omega}\\
		\dot{\Upsilon}
	\end{bmatrix}=\begin{bmatrix}\frac{1}{2}\Xi(\omega)q\\
		v\\
		R(q)\frac{u(1)}{m_{s}}E_{z}-gE_{z}\\
		\mathcal{J}^{-1}\big(u(2:4)-\omega\times\mathcal{J}\omega\big)\\
		A\big(-\Upsilon+G+Wu-\Gamma\big)
	\end{bmatrix}.\label{equ:right_hand_side}
\end{equation}
To facilitate estimation ,the augmented dynamics in \eqref{equ:augmented_dynamics} can be restructured into a geometric form, consistent with the methodologies in \cite{hashim2021geometric,ghanizadegan2024quaternion}:
\[
\dot{x}=\operatorname{f}^{c}(q,r,\omega,u)x
\]
\small{}
\begin{equation}
	\begin{bmatrix}\dot{q}\\
		\dot{r}\\
		\dot{v}\\
		\dot{\omega}\\
		\dot{\Upsilon}\\
		0
	\end{bmatrix}=\begin{bmatrix}\frac{1}{2}\Xi(\omega) & 0 & 0 & 0 & 0\\
		0 & 0 & \mathbf{I}_{3} & 0 & 0\\
		0 & 0 & 0 & 0 & R(q)\frac{u(1)}{m_{s}}E_{z}-gE_{z}\\
		0 & 0 & 0 & 0 & \mathcal{J}^{-1}\big(u(2:4)-\omega\times\mathcal{J}\omega\big)\\
		0 & 0 & 0 & 0 & A\big(-\Upsilon+G+Wu-\Gamma\big)\\
		0 & 0 & 0 & 0 & 0
	\end{bmatrix}\begin{bmatrix}q\\
		r\\
		v\\
		\omega\\
		\Upsilon\\
		1
	\end{bmatrix},\label{con_trans_fun_equ}
\end{equation}
where the state is extended to $x\in\mathbb{R}^{20}$ for geometric reconstruction. This continuous system in \eqref{con_trans_fun_equ} can be discretized over a sampling interval $T$ using the matrix exponential:
\begin{equation}
	\begin{bmatrix}q_{k}\\
		r_{k}\\
		v_{k}\\
		\omega_{k}\\
		\Upsilon_{k}\\
		1
	\end{bmatrix}=\exp(\operatorname{f}_{k-1}^{c}T)\begin{bmatrix}q_{k-1}\\
		r_{k-1}\\
		v_{k-1}\\
		\omega_{k-1}\\
		\Upsilon_{k-1}\\
		1
	\end{bmatrix},\label{dis_trans_fun_equ}
\end{equation}
where $\operatorname{f}_{k-1}^{c}=\operatorname{f}^{c}(q_{k-1},r_{k-1},\omega_{k-1},u_{k-1})$  and $x_{k}=[q_{k},r_{k},v_{k},\omega_{k},\Upsilon_{k},1]^{\top}$ represents the discrete state vector. The measured signals: position $r_{m_{k}}$, orientation
$q_{m_{k}}$, and angular velocity $\omega_{m_{k}}$ are affected by additive noise, as described in \eqref{process_measure_noise_mat_eaq}. Given the assumption that process noise remains constant over $T$, the continuous covariance is discretized as $\mathsf{Q}_{k}=\mathsf{Q}T$.

\subsection{Unscented Transformation  (UT)}

The Unscented Transformation (UT) is the core mechanism of the UKF, allowing for the precise propagation of the mean $\bar{x}$ and covariance $P_{x}$ of a random variable $x$ through the nonlinear functions. By utilizing a deterministic set of sigma points, the UT avoids the linearization errors inherent in the EKF. The selection of these points is governed by three scaling parameters:
\begin{enumerate}
	\item Primary scaling $(\varphi)$: Controls spread of sigma points ($10^{-4}$ to $1$). Smaller
	values produce tighter clusters; larger values give a wider spread. 
	\item Secondary scaling $(\gamma)$: Incorporates prior distribution knowledge (optimally $\gamma=2$ for Gaussian distributions).
	\item Tertiary scaling $(\sigma)$: Typically set to 0. 
\end{enumerate}
The weights for the mean ($\mu^{m}$) and covariance ($\mu^{c}$), along with an additional scaling factor $\eta$, are defined as \cite{rhudy2013understanding}:
\begin{equation}
	\begin{aligned}\eta & =\varphi^{2}(N+\sigma)-N,\\
		\mu_{0}^{m} & =\eta/(N+\eta),\\
		\mu_{0}^{c} & =\eta/(N+\eta)+1-\varphi^{2}+\gamma,\\
		\mu_{i}^{m} & =\mu_{i}^{c}=1/[2(N+\eta)],\quad i=1,...,2N,
	\end{aligned}
	\label{scaling_para_equ}
\end{equation}
where $N$ is the state dimension. The $2N+1$ sigma points are generated using the Cholesky factor of the covariance matrix:
\begin{equation}
	\mathcal{L}=\begin{bmatrix}\bar{x},\bar{x}+\sqrt{N+\eta}\sqrt{P_{x}},\bar{x}-\sqrt{N+\eta}\sqrt{P_{x}}\end{bmatrix},\label{segma_points_equ}
\end{equation}
Passing these points through the nonlinear function $\operatorname{f}$ yields the transformed set:
\begin{equation}
	\mathcal{S}^{(i)}=\operatorname{f}(\mathcal{L}^{(i)}),\qquad i=0,1,...,2N.
\end{equation}
Finally, the posterior mean and covariance are reconstructed via weighted averaging:
\begin{equation}
	\begin{aligned}\bar{y} & \approx\sum_{i=0}^{2N}\mu_{i}^{m}\mathcal{S}^{(i)},\\
		P_{y} & \approx\sum_{i=0}^{2N}\mu_{i}^{c}(\mathcal{S}^{(i)}-\bar{y})(\mathcal{S}^{(i)}-\bar{y})^{\top}.
	\end{aligned}
	\label{mean_cova_equ}
\end{equation}

\subsection{QUKF Implementation}

The QUKF recursion begins with $\hat{x}_{0}$ and
$P_{0}$. Unlike conventional UKF, quaternions require special handling,
since they cannot be directly added/subtracted (see Section \ref{quat_subsection},
\eqref{sum_sub_equ}, \eqref{Quat_sub_equ}), and must satisfy the
unit norm constraint (four elements, three DoF) which requires covariance
adjustments. Thus, the initial state and covariance are defined as:
\begin{equation}
	\begin{aligned}\hat{x}_{0} & =\begin{bmatrix}\hat{x}_{0,q}^{\top},\hat{x}_{0,nq}^{\top}\end{bmatrix}^{\top}\in\mathbb{R}^{N},\\
		P_{0} & =\operatorname{blkdiag}\big(P_{0,q},P_{0,nq}\big)\in\mathbb{R}^{(N-1)\times(N-1)},
	\end{aligned}
	\label{initail_x_P_equ}
\end{equation}
with $\hat{x}_{0,q}\in\mathbb{S}^{3}$ represents the quaternion term
with its covariance $P_{0,q}\in\mathbb{R}^{3\times3}$ that satisfies
the unit constraint and $\hat{x}_{0,nq}\in\mathbb{R}^{N-4}$ represents
the non-quaternion terms with covariance $P_{0,nq}\in\mathbb{R}^{(N-4)\times(N-4)}$.
At each time step $k$, sigma points are generated from the prior
state estimate $\hat{x}_{k-1}\in\mathbb{R}^{N}$ and covariance $P_{k-1}\in\mathbb{R}^{(N-1)\times(N-1)}$
using a modified version of equation \eqref{segma_points_equ} as:
\begin{align}
	\mathcal{L}_{k-1}= & \left[\hat{x}_{k-1},\hat{x}_{k-1}+\sqrt{N-1+\eta}\sqrt{P_{k-1}},\right.\nonumber \\
	& \hspace{5em}\left.\hat{x}_{k-1}-\sqrt{N-1+\eta}\sqrt{P_{k-1}}\right].\label{chi_segma_points_equ}
\end{align}
To handle the addition and subtraction operations of the quaternions
and rotation vectors in \eqref{chi_segma_points_equ}, let
\begin{align*}
	\hat{C}_{k-1} & =\sqrt{N-1+\eta}\sqrt{P_{k-1}}\in\mathbb{R}^{(N-1)\times(N-1)},\\
	\hat{c}_{k-1}^{(i)} & =\sqrt{N-1+\eta}\sqrt{P_{k-1}}\in\mathbb{R}^{(N-1)}~(i^{th}~\text{column of}~\hat{C}_{k-1})
\end{align*}
divide $\hat{c}_{k-1}^{(i)}$ and $\hat{x}_{k-1}$ into their orientation
and non-orientation components, as it was done in \eqref{initail_x_P_equ}:
\begin{equation}
	\begin{aligned}\hat{x}_{k-1} & =\begin{bmatrix}\hat{x}_{k-1,q}^{\top},\hat{x}_{k-1,nq}^{\top}\end{bmatrix}^{\top}\in\mathbb{R}^{N},\\
		\hat{c}_{k-1}^{(i)} & =\begin{bmatrix}\hat{c}_{k-1,r}^{(i)\top},\hat{c}_{k-1,nr}^{(i)\top}\end{bmatrix}^{\top}\in\mathbb{R}^{N-1}.
	\end{aligned}
	\label{kth_x_P_equ}
\end{equation}
utilizing the custom addition and subtraction operators in \eqref{sum_sub_equ}
to map from ($\mathbb{S}^{3}\times\mathbb{R}^{3}\rightarrow\mathbb{S}^{3}$),
equation \eqref{chi_segma_points_equ} is modified to calculate the
sigma points as follows:
\begin{equation}
	\mathcal{L}_{k-1}=\begin{bmatrix}\hat{x}_{k-1},\hat{x}_{k-1}\oplus\hat{C}_{k-1},\hat{x}_{k-1}\ominus\hat{C}_{k-1}\end{bmatrix}\in\mathbb{R}^{N\times(2N-1)},\label{sigma_points_operators_equ}
\end{equation}
where the addition and subtraction expressions in \eqref{sigma_points_operators_equ}
are given as follows:
\begin{align*}
	\hat{x}_{k-1}\oplus\hat{C}_{k-1} & \equiv\begin{bmatrix}\hat{x}_{k-1,q}\oplus\hat{c}_{k-1,r}^{(i)}\\
		\hat{x}_{k-1,nq}+\hat{c}_{k-1,nr}^{(i)}
	\end{bmatrix}\in\mathbb{R}^{(N\times N-1)},\\
	\hat{x}_{k-1}\ominus\hat{C}_{k-1} & \equiv\begin{bmatrix}\hat{x}_{k-1,q}\ominus\hat{c}_{k-1,r}^{(i)}\\
		\hat{x}_{k-1,nq}-\hat{c}_{k-1,nr}^{(i)}
	\end{bmatrix}\in\mathbb{R}^{(N\times N-1)}.
\end{align*}
Sigma points in \eqref{sigma_points_operators_equ} are propagated
through the nonlinear state transition function, $\operatorname{f}$
in \eqref{dis_trans_fun_equ} as:
\begin{equation}
	\mathcal{L}_{k|k-1}^{(i)}=\operatorname{f}(\mathcal{L}_{k-1}^{(i)},u_{k-1}),\hspace{1em}i=0,1,...,2(N-1),\label{chi_nonlinear_tran_fun_equ}
\end{equation}
with process noise excluded (zero-mean additive). The weight vectors
in \eqref{scaling_para_equ} and the weighted averaging in \eqref{mean_cova_equ}
also adapt to reflect the reduced dimensionality as: 
\begin{equation}
	\begin{aligned}\eta & =\varphi^{2}((N-1)+\sigma)-(N-1),\\
		\mu_{0}^{m} & =\eta/((N-1)+\eta),\\
		\mu_{0}^{c} & =\eta/((N-1)+\eta)+1-\varphi^{2}+\gamma,\\
		\mu_{i}^{m} & =\mu_{i}^{c}=1/(2((N-1)+\eta)),\hspace{1em}i=1,...,2(N-1),
	\end{aligned}
	\label{modified_scaling_para_equ}
\end{equation}
\begin{align}
	\hat{x}_{k|k-1}= & \sum_{i=0}^{2(N-1)}\mu_{i}^{m}\mathcal{L}_{k|k-1}^{(i)},\nonumber \\
	P_{k|k-1}= & {\displaystyle \sum_{i=0}^{2(N-1)}\mu_{i}^{c}\big(\mathcal{L}_{k|k-1}^{(i)}}-\hat{x}_{k|k-1}\big)\big(\mathcal{L}_{k|k-1}^{(i)}-\hat{x}_{k|k-1}\big)^{\top}\nonumber \\
	& +\mathsf{Q}_{k-1}.\label{sigma_mean_cova_equ}
\end{align}
The process noise covariance $\mathsf{Q}_{k}$ is added to the error covariance under the assumption of additive noise. For quaternion components in the unscented filtering framework, the weighted mean of the sigma points in \eqref{sigma_mean_cova_equ} must be computed using the quaternion averaging method in \eqref{weighted_avrg_equ}, since standard vector averaging is not valid on $\mathbb{S}^{3}$, followed by normalization to enforce the unit-norm constraint. This approach provides a computationally efficient approximation that is well suited for real-time applications with moderate attitude dispersion. Alternative methods, such as the Riemannian (Karcher) mean or averaging performed in the Lie algebra via logarithmic and exponential mappings \cite{angulo2014riemannian, voight2021quaternion}, can offer improved accuracy when quaternion dispersion is large, at the cost of increased computational complexity. Given the real-time constraints and the bounded attitude variations typical of assistive aerial transportation tasks, the adopted averaging scheme represents a practical trade-off between numerical stability and computational efficiency. Likewise, subtraction of quaternion components requires the operator in \eqref{Quat_sub_equ}, which maps $(\mathbb{S}^{3}\times\mathbb{S}^{3}\rightarrow\mathbb{R}^{3})$ and replaces ordinary vector subtraction.

To formalize the process, decompose both the predicted
state $\hat{x}_{k|k-1}$ and any sigma point $\mathcal{L}_{k|k-1}^{(i)}$
into their quaternion and non-quaternion components, consistent with
the earlier formulation as:
\begin{equation}
	\begin{aligned}\hat{x}_{k|k-1} & =\begin{bmatrix}\hat{x}_{k|k-1,q}^{\top},\hat{x}_{k|k-1,nq}^{\top}\end{bmatrix}^{\top}\in\mathbb{R}^{N},\\
		\mathcal{L}_{k|k-1}^{(i)} & =\begin{bmatrix}\mathcal{L}_{k|k-1,q}^{(i)\top},\mathcal{L}_{k|k-1,nq}^{(i)\top}\end{bmatrix}^{\top}\in\mathbb{R}^{N}.
	\end{aligned}
	\label{kth_x_sigma_equ}
\end{equation}
Based on \eqref{kth_x_sigma_equ}, the expression
in \eqref{sigma_mean_cova_equ} is modified as follows:
\begin{align}
	\hat{x}_{k|k-1}= & \begin{bmatrix}\operatorname{WAve}(\{\mathcal{L}_{k|k-1,q}^{(i)}\},\{\mu_{i}^{m}\})\\
		\sum_{i=0}^{2(N-1)}\mu_{i}^{m}\mathcal{L}_{k|k-1,nq}^{(i)}
	\end{bmatrix}\in\mathbb{R}^{N},\nonumber \\
	P_{k|k-1}= & {\displaystyle \sum_{i=0}^{2(N-1)}\mu_{i}^{c}\big(\mathcal{L}_{k|k-1}^{(i)}}\ominus\hat{x}_{k|k-1}\big)\big(\mathcal{L}_{k|k-1}^{(i)}\ominus\hat{x}_{k|k-1}\big)^{\top}\nonumber \\
	& +\mathsf{Q}_{k-1},\label{quat_sigma_mean_cova_equ}
\end{align}
where the $(\mathcal{L}_{k|k-1}^{(i))}\ominus\hat{x}_{k|k-1})$ in
\eqref{quat_sigma_mean_cova_equ} is given as:
\begin{equation}
	\mathcal{L}_{k|k-1}^{(i)}\ominus\hat{x}_{k|k-1}\equiv\begin{bmatrix}\mathcal{L}_{k|k-1,q}^{(i)}\ominus\hat{x}_{k|k-1,q}\\
		\mathcal{L}_{k|k-1,nq}^{(i)}-\hat{x}_{k|k-1,nq}
	\end{bmatrix}\in\mathbb{R}^{N-1}.\label{equivalent_process_equ}
\end{equation}
These values correspond to the predicted mean and
covariance at time step $k$, commonly referred to as the a priori
state and covariance estimates. The transformed sigma points are then
passed through the observation function. As in the prediction step,
measurement noise is omitted from the function since it is assumed
additive and zero-mean:
\begin{equation}
	\mathcal{Q}_{k|k-1}^{(i)}=\operatorname{h}(\mathcal{L}_{k|k-1}^{(i)}),\hspace{1em}i=0,1,...,2(N-1),\label{chi_non_measure_fun_equ}
\end{equation}
where $\mathcal{Q}\in\mathbb{R}^{m\times2(N-1)}$ collects the output
sigma points and $m$ is the output dimension. These sigma points
are then used to compute the predicted output, the output covariance,
and the state-output cross-covariance. All calculations follow the
same quaternion operations introduced in \eqref{kth_x_sigma_equ},
\eqref{quat_sigma_mean_cova_equ}, and \eqref{equivalent_process_equ}:
\begin{equation}
	\begin{aligned}\hat{y}_{k|k-1}= & \begin{bmatrix}\operatorname{WAve}(\{\mathcal{Q}_{k|k-1,q}^{(i)}\},\{\mu_{i}^{m}\})\\
			\sum_{i=0}^{2(N-1)}\mu_{i}^{m}\mathcal{Q}_{k|k-1,nq}^{(i)}
		\end{bmatrix},\\
		P_{k}^{xy}= & \sum_{i=0}^{2(N-1)}\mu_{i}^{c}\big(\mathcal{L}_{k|k-1}^{(i)}\ominus\hat{x}_{k|k-1}\big)\big(\mathcal{Q}_{k|k-1}^{(i)}\ominus\hat{y}_{k|k-1}\big)^{\top},\\
		P_{k}^{yy}= & \sum_{i=0}^{2(N-1)}\mu_{i}^{c}\big(\mathcal{Q}_{k|k-1}^{(i)}\ominus\hat{y}_{k|k-1}\big)\big(\mathcal{Q}_{k|k-1}^{(i)}\ominus\hat{y}_{k|k-1}\big)^{\top}\\
		& +\mathsf{R}_{k}.
	\end{aligned}
	\label{output_mean_cova_equ}
\end{equation}
The measurement noise covariance $\mathsf{R}$ is
incorporated into the output covariance matrix in \eqref{output_mean_cova_equ}
under the additive noise assumption. These matrices are then used
to compute the Kalman gain as:
\begin{equation}
	K_{k}=P_{k}^{xy}\big(P_{k}^{yy}\big)^{-1}.\label{kalman_gain_equ}
\end{equation}
The gain $K_{k}$ updates both the state and covariance
estimates as: 
\begin{equation}
	\begin{aligned}\hat{x}_{k} & =\hat{x}_{k|k-1}\oplus K_{k}\big(y_{k}\ominus\hat{y}_{k|k-1}\big),\\
		P_{k} & =P_{k|k-1}-K_{k}P_{k}^{yy}K_{k}^{\top},
	\end{aligned}
	\label{final_est_cova_equ}
\end{equation}
where $y_{k}\in\mathbb{R}^{m}$ is the measurement
vector (noisy position, quaternion, and angular velocity). The residual
requires quaternion-aware operations such that:
\begin{equation}
	y_{k}\ominus\hat{y}_{k|k-1}\equiv\begin{bmatrix}y_{k,q}\ominus\hat{y}_{k|k-1,q}\\
		y_{k,nq}-\hat{y}_{k|k-1,nq}
	\end{bmatrix}.\label{ominus_process_equ}
\end{equation}
and the correction is applied as
\begin{equation}
	\hat{x}_{k|k-1}\oplus r_{k|k-1}\equiv\begin{bmatrix}\hat{x}_{k|k-1,q}\oplus r_{k|k-1,q}\\
		\hat{x}_{k|k-1,nq}+r_{k|k-1,nq}
	\end{bmatrix}.\label{oplus_process_equ}
\end{equation}
where $r_{k|k-1}=K_{k}\big(y_{k}\ominus\hat{y}_{k|k-1}\big)$.
Here, $\hat{x}_{k}$ and $P_{k}$ denote the posterior state and covariance
estimates, which serve as priors for the next recursion step of the
QUKF. The QUKF implementation steps are outlined in Algorithm \ref{QUKF_alg}.

\begin{algorithm}[h!]
	{ \small
		\caption{\label{QUKF_alg} Quaternion-based Unscented Kalman Filter}
		\textbf{Initialization}: 
		\begin{enumerate}
			\item[{\footnotesize{}{}{}{}{}1:}] Select the scaling parameters $\varphi$, $\gamma$, $\sigma$, and
			$\eta\in\mathbb{R}$ along with the state vector length $N$. 
			\item[{\footnotesize{}{}{}{}{}2:}] Calculate the weight vectors $\mu^{m}$ and $\mu^{c}$, as expressed
			in \eqref{modified_scaling_para_equ}. 
			\item[{\footnotesize{}{}{}{}{}3:}] Define process and measurement noise covariance matrices, $\mathsf{Q}$
			and $\mathsf{R}$, see \eqref{process_measure_noise_mat_eaq}. 
			\item[{\footnotesize{}{}{}{}{}4:}] Set $k=1$ and initialize $\hat{x}_{0}\in\mathbb{R}^{N}$, and $P_{0}\in\mathbb{R}^{(N-1)\times(N-1)}$,
			see \eqref{initail_x_P_equ}. 
		\end{enumerate}
		\textbf{While $\boldsymbol{y_{k}}$ data exists} 
		\begin{enumerate}
			\item[{\footnotesize{}{}{}{}{}5:}] Calculate the sigma points, as: 
			\item[] $\mathcal{L}_{k-1}=\big[\hat{x}_{k-1},\hat{x}_{k-1}\oplus\hat{C}_{k-1},\hat{x}_{k-1}\ominus\hat{C}_{k-1}\big]$,
			see \eqref{sigma_points_operators_equ}. 
			\item[] /{*} Prediction {*}/ 
			\item[{\footnotesize{}{}{}{}{}6:}] Propagate each sigma point through prediction, as: 
			\item[] \textbf{for} $i=\{0,1\ldots,2(N-1)\}$ 
			\item[] $\qquad\mathcal{L}_{k|k-1}^{(i)}=\operatorname{f}(\mathcal{L}_{k-1}^{(i)},u_{k-1})$. 
			\item[] \textbf{end for},\textbf{ }see \eqref{chi_nonlinear_tran_fun_equ}. 
			\item[{\footnotesize{}{}{}{}{}7:}] Calculate mean and covariance of predicted state, as: 
			\item[] $\hat{x}_{k|k-1}=\begin{bmatrix}\operatorname{WAve}(\{\mathcal{L}_{k|k-1,q}^{(i)}\},\{\mu_{i}^{m}\})\\
				\sum_{i=0}^{2N}\mu_{i}^{m}\mathcal{L}_{k|k-1,nq}^{(i)}
			\end{bmatrix}$. 
			\item[] $P_{k|k-1}=\mathsf{Q}_{k-1}+\sum_{i=0}^{2N}\mu_{i}^{c}(\mathcal{L}_{k|k-1}^{(i)}\ominus\hat{x}_{k|k-1})(\mathcal{L}_{k|k-1}^{(i)}\ominus\hat{x}_{k|k-1})^{\top}$,\textbf{
			}see \eqref{quat_sigma_mean_cova_equ}. 
			\item[] /{*} Observation {*}/ 
			\item[{\footnotesize{}{}{}{}{}8:}] Propagate each sigma point through observation, as: 
			\item[] \textbf{for} $i=\{0,1\ldots,2(N-1)\}$ 
			\item[] $\qquad\mathcal{Q}_{k|k-1}^{(i)}=\operatorname{h}(\mathcal{L}_{k|k-1}^{(i)})$ 
			\item[] \textbf{end for},\textbf{ }see \eqref{chi_non_measure_fun_equ}. 
			\item[{\footnotesize{}{}{}{}{}9:}] Calculate mean and covariance of predicted output, as: 
			\item[] $\hat{y}_{k|k-1}=\begin{bmatrix}\operatorname{WAve}(\{\mathcal{Q}_{k|k-1,q}^{(i)}\},\{\mu_{i}^{m}\})\\
				\sum_{i=0}^{2N}\mu_{i}^{m}\mathcal{Q}_{k|k-1,nq}^{(i)}
			\end{bmatrix}$ 
			\item[] $P_{k}^{yy}=\mathsf{R}_{k}+\sum_{i=0}^{2N}\mu_{i}^{c}\big(\mathcal{Q}_{k|k-1}^{(i)}\ominus\hat{y}_{k|k-1}\big)\big(\mathcal{Q}_{k|k-1}^{(i)}\ominus\hat{y}_{k|k-1}\big)^{\top}$,
			see \eqref{output_mean_cova_equ}. 
			\item[{\footnotesize{}{}{}{}{}10:}] Calculate cross-covariance of state and output, as: 
			\item[] $P_{k}^{xy}=\sum_{i=0}^{2N}\mu_{i}^{c}\big(\mathcal{L}_{k|k-1}^{(i)}\ominus\hat{x}_{k|k-1}\big)\big(\mathcal{Q}_{k|k-1}^{(i)}\ominus\hat{y}_{k|k-1}\big)^{\top}$,
			see \eqref{output_mean_cova_equ}. 
			\item[] /{*} Update {*}/ 
			\item[{\footnotesize{}{}{}{}{}11:}] Calculate Kalman gain $K$, as: 
			\item[] $k_{k}=P_{k}^{xy}(P_{k}^{yy})^{-1}$, see \eqref{kalman_gain_equ}. 
			\item[{\footnotesize{}{}{}{}{}12:}] Update state estimate vector and error covariance matrix, as: 
			\item[] $\hat{x}_{k}=\hat{x}_{k|k-1}\oplus K_{k}\big(y_{k}\ominus\hat{y}_{k|k-1}\big)$ 
			\item[] $P_{k}=P_{k|k-1}-K_{k}P_{k}^{yy}K_{k}^{\top}$, see \eqref{final_est_cova_equ}. 
			\item[{\footnotesize{}{}{}{}{}13:}] $k=k+1$. 
		\end{enumerate}
		\textbf{end while} 
	}
\end{algorithm}

\section{Numerical Results\label{results-discussion}}
\subsection{Simulation Setup}

The performance of the human-UAV cooperative payload transportation
system with the proposed Quaternion-based Unscented Kalman Filter (QUKF)
for external wrench estimation was evaluated by extensive simulations.
This section presents the system setup, estimator parameters, and
analysis of results. The cooperative system was modeled as a single
rigid body with the following parameters: total mass (UAVs + payload)
$m_{s}=3.49$ kg, inertia matrix $\mathcal{J}{s}=\text{diag}(3.227,0.061,3.277)$
kg·m$^{2}$, gravitational acceleration $g=9.81$ m/s$^{2}$, payload
length $\mathfrak{L}=2$ m, maximum allowable UAV thrust $u_{max}=35$
N, and integration step $T=0.01$ s. The admittance controller was
tuned to provide intuitive human interaction using a virtual mass
$\mathbb{M}_{v}=\text{diag}(1,1,1)$ kg, virtual damping $\mathbb{C}_{v}=\text{diag}(1.59,1.59,1.59)$
N·s/m, and zero stiffness $\mathbb{K}_{v}=0$ N/m.

The QUKF was implemented with the parameter set in Table \ref{tab:qukf_parameters}.
The quaternion-based formulation ensured robustness against singularities
in Euler angle representations and eliminated long-term kinematic
drift. The unit quaternion constraint was enforced via normalization
at each update step.

\begin{table}[ht]
	\centering
	\caption{\label{tab:qukf_parameters}QUKF Parameters.}
	\begin{tabular}{lll}
		\hline 
		Symbol  & Definition  & Value\tabularnewline
		\hline 
		\hline 
		$\mathsf{Q}_k$  & Process noise  & $\text{diag}(10^{-4}\mathbf{I}_{3},\rightarrow$ Rotation vector\tabularnewline
		& covariance  & $\hspace{0.6cm}10^{-4}\mathbf{I}_{3},\rightarrow$ Position \tabularnewline
		&  & $\hspace{0.6cm}10^{-1}\mathbf{I}_{3},\rightarrow$ Velocity\tabularnewline
		&  & $\hspace{0.6cm}10^{-3}\mathbf{I}_{3},\rightarrow$ Angular velocity\tabularnewline
		&  & $\hspace{0.6cm}10^{-2}\mathbf{I}_{6},\rightarrow$ Ext. force/torque\tabularnewline
		&  & $\hspace{1.3cm}0)\rightarrow$ Dummy state\tabularnewline
		\hline 
		$\mathsf{R}$  & Measurement  & $\text{diag}(10^{-4}\mathbf{I}_{3},\rightarrow$ Rotation vector\tabularnewline
		& noise covariance  & $\hspace{0.6cm}10^{-4}\mathbf{I}_{3},\rightarrow$ Position \tabularnewline
		&  & $\hspace{0.6cm}10^{-3}\mathbf{I}_{3})\rightarrow$ Angular velocity\tabularnewline
		\hline 
		$P_{0}$  & Initial covariance  & $\text{diag}(10^{-4}\mathbf{I}_{3},\rightarrow$ Rotation vector\tabularnewline
		&  & $\hspace{0.6cm}10^{-2}\mathbf{I}_{3},\rightarrow$ Position \tabularnewline
		&  & $\hspace{0.6cm}10^{-2}\mathbf{I}_{3},\rightarrow$ Velocity\tabularnewline
		&  & $\hspace{0.6cm}10^{-2}\mathbf{I}_{3},\rightarrow$ Angular velocity\tabularnewline
		&  & $\hspace{0.6cm}10^{0}~~\mathbf{I}_{6},\rightarrow$ Ext. force/torque\tabularnewline
		&  & $\hspace{1.3cm}0)\rightarrow$ Dummy state\tabularnewline
		\hline 
		$N$  & State vector dim.  & $\hspace{0.8cm}$20\tabularnewline
		$m$  & Output vector dim.  & $\hspace{0.8cm}$10\tabularnewline
		$\delta$  & Positive constant  & $\hspace{0.8cm}$72\tabularnewline
		$\varphi$  & Scaling parameter  & $\hspace{0.8cm}$1\tabularnewline
		$\gamma$  & Scaling parameter  & $\hspace{0.8cm}$2\tabularnewline
		$\sigma$  & Scaling parameter  & $\hspace{0.8cm}$0\tabularnewline
		$\eta$  & Scaling parameter  & $\hspace{0.8cm}$0\tabularnewline
		\hline 
	\end{tabular}
\end{table}

The simulation was carried out in MATLAB R2024b on a laptop with an Intel i7-1355U CPU (1.70 GHz) and 16 GB RAM. The computational cost of QUKF is dominated by sigma point generation/propagation, covariance reconstruction, and Kalman gain computation, all scaling with the state dimension $N$. For $N=20$ and measurement dimension $m=10$, the main operations scale as: sigma point propagation $O(N^3)$, covariance reconstruction $O(N^3)$, Kalman gain computation $O(m^3+Nm^2)$, and covariance update $O(N^2m)$. Thus, the overall complexity remains $O(N^3)$, consistent with standard UKF formulations. However, by reducing the quaternion dimension from $N$ to $(N-1)$, the number of propagated sigma points decreases from $(2N+1)$ to $(2(N-1)+1)$, which lowers the constant computation overhead. The implementation achieved an average runtime of 0.7656 ms per update, well below the 10 ms requirement for real-time $100$ Hz operation. Scaling tests further validated the expected $O(N^3)$ trend.

\subsection{Human-UAV Interaction Scenario}

\begin{figure}[b!]
	\centering
	\includegraphics[clip,width=0.95\columnwidth]{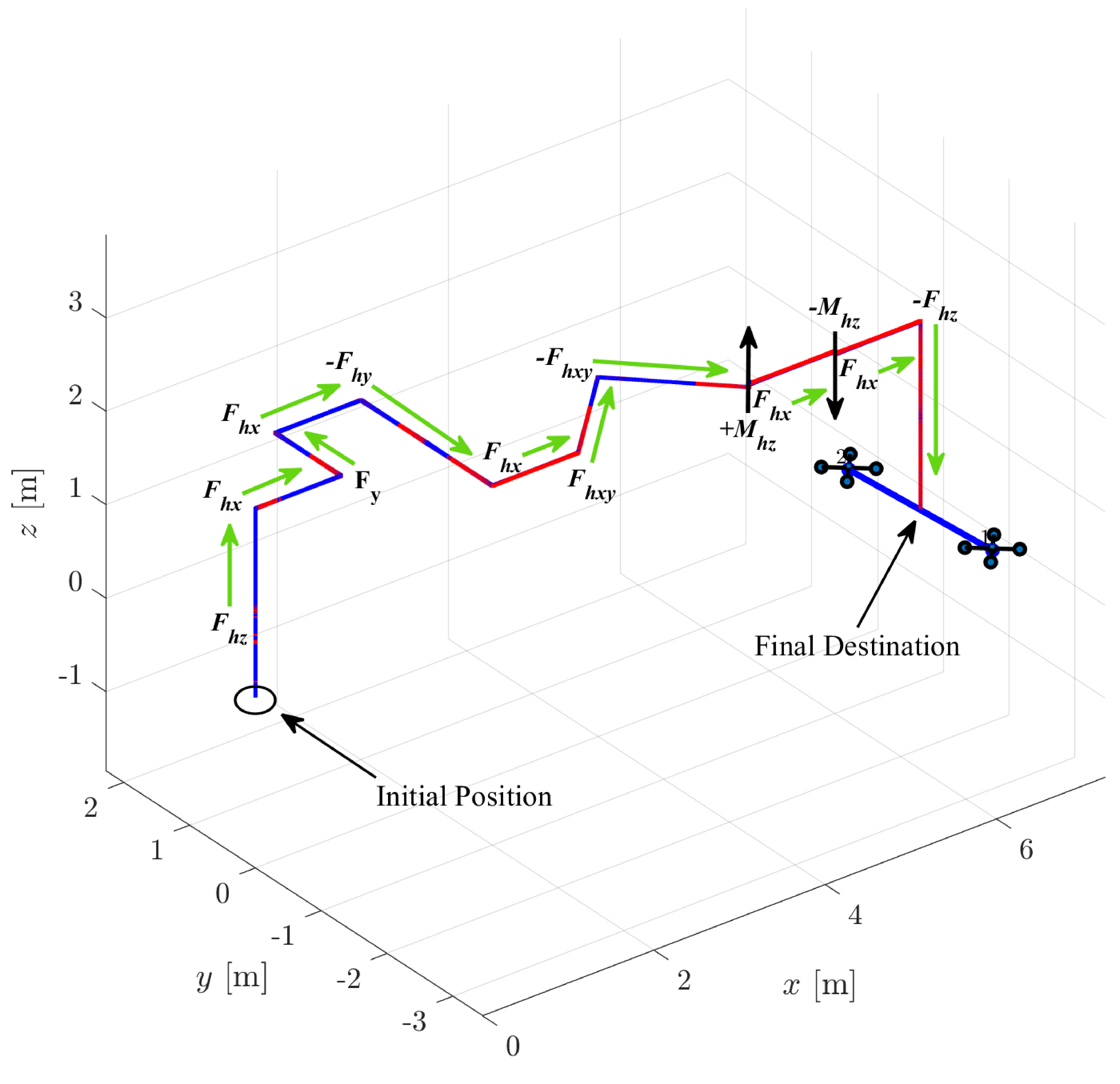}
	\caption{\label{Fig3:main_matlab}Trajectory of the assistive system under human guidance in 3D space, with red arrows indicating applied forces.}
\end{figure}

To evaluate cooperative performance, a human-UAV interaction scenario was implemented. The system was initialized to hover at the starting position. A human operator then applied a sequence of forces and torques to the payload in multiple directions, guiding the system along a three-dimensional trajectory toward a designated destination, as shown in Figure \ref{Fig3:main_matlab}.

The proposed QUKF played a central role in enabling this interaction. By accurately estimating the external wrench in real time, the filter allowed the admittance controller to correctly interpret the operator’s input and generate smooth reference trajectories. The results show that the system responded effectively to human guidance, exhibiting stable attitude regulation and seamless transitions between motion states. The visualization confirms that the UAVs followed the applied forces while maintaining robust stability, highlighting the importance of reliable wrench estimation for a safe and intuitive physical human-UAV interaction.

\subsection{Results Discussion}

\begin{figure*}[h!]
	\centering
	\includegraphics[clip,width=2\columnwidth]{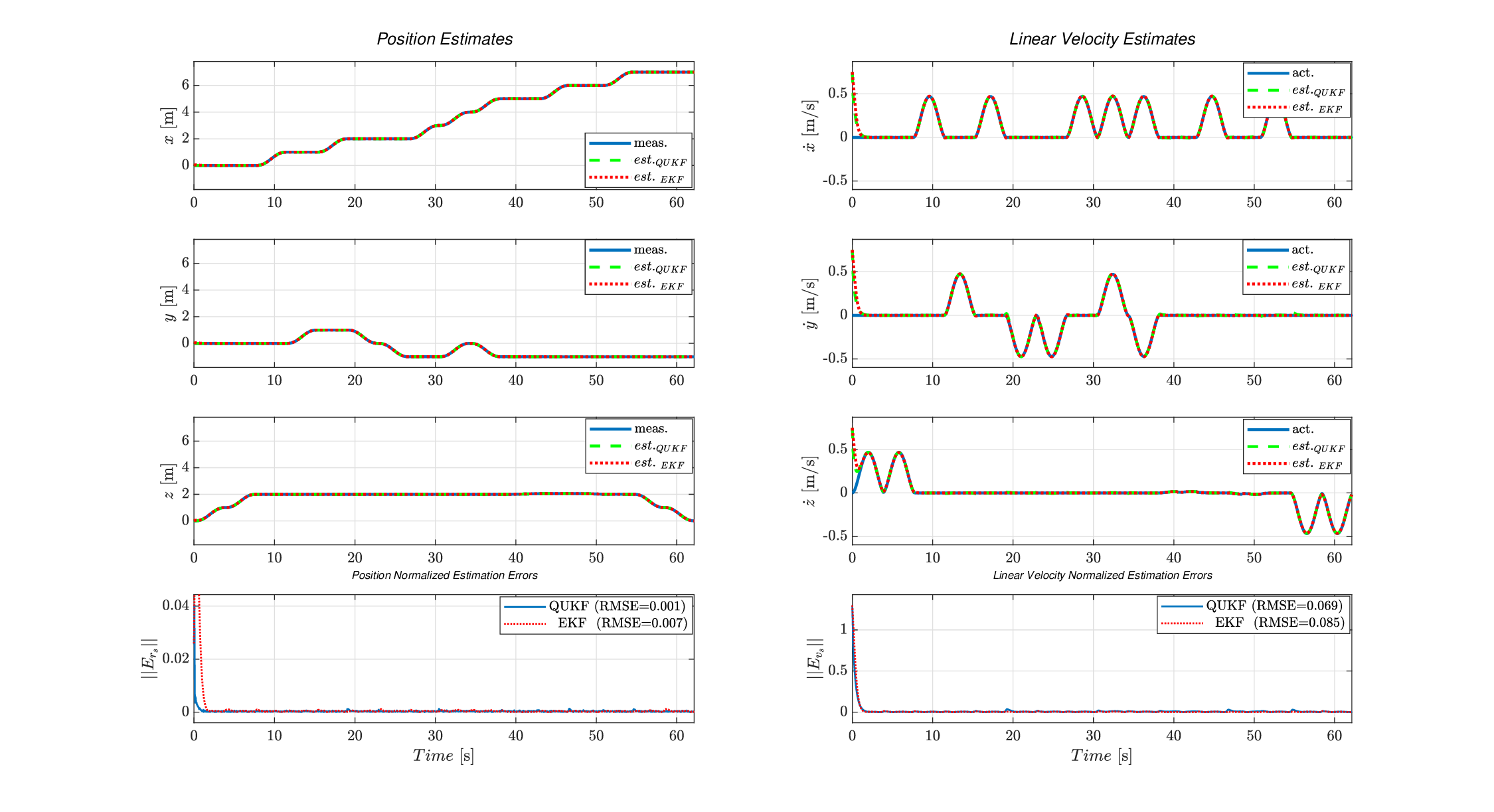}
	\caption{\label{Fig4:pos_vel_estimation}Position and linear velocity estimation with normalized errors: A performance comparison between the proposed QUKF and baseline EKF.}
\end{figure*}

\begin{figure*}[h!]
	\centering
	\includegraphics[clip,width=2\columnwidth]{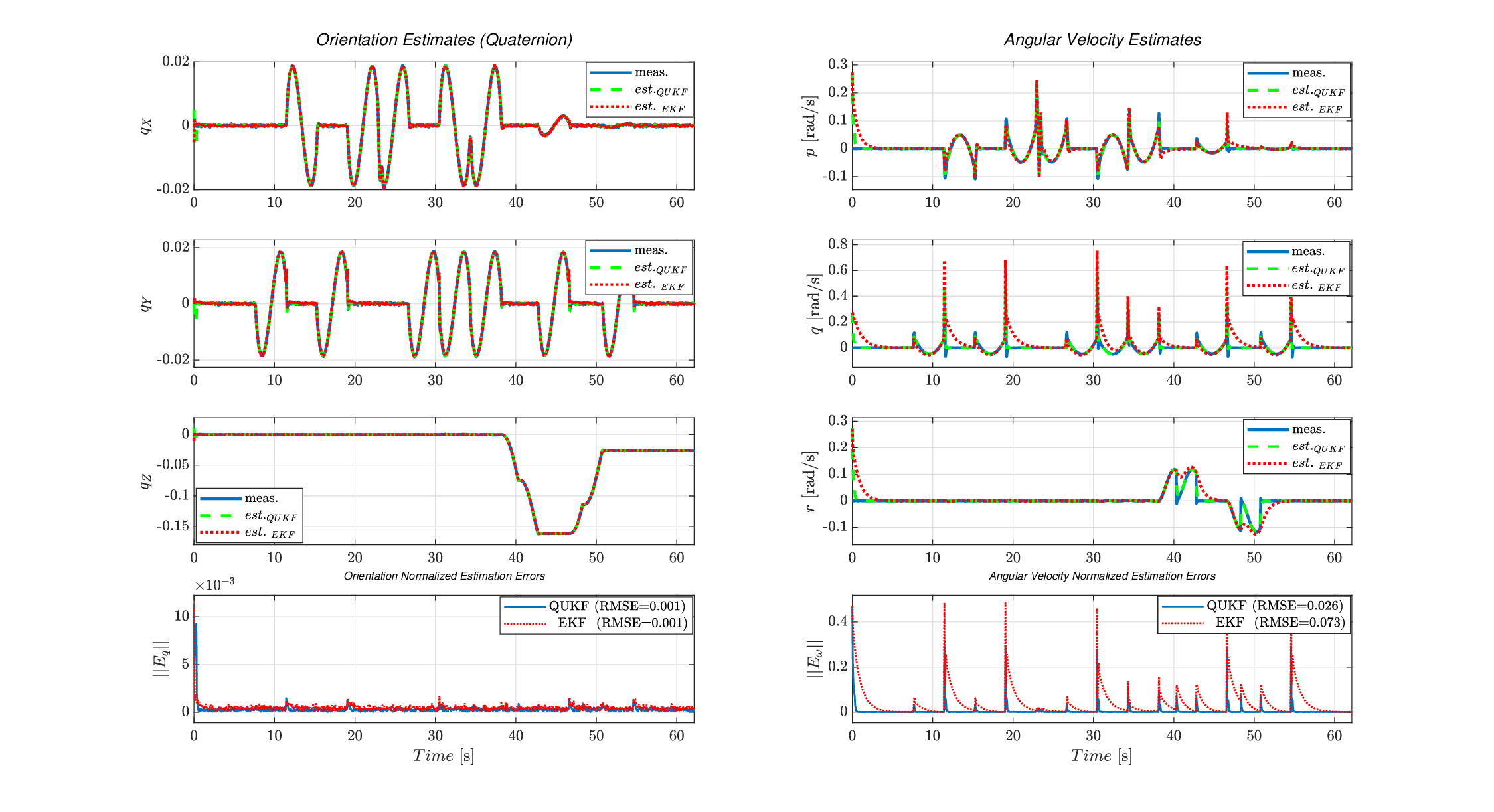}
	\caption{\label{Fig5_quat_orient_vel_estimation}Orientation and angular velocity estimation with normalized errors: A performance comparison between the proposed QUKF and baseline EKF.}
\end{figure*}

The state estimation results are presented in Figures \ref{Fig4:pos_vel_estimation} and \ref{Fig5_quat_orient_vel_estimation}, which show the actual (desired), measured, and estimated states, including position, linear velocity, orientation (quaternion), and angular velocity, along with their normalized estimation errors.
Figure \ref{Fig4:pos_vel_estimation} shows that the QUKF precisely estimated the position and linear velocity, rapidly reducing estimation errors to near zero and maintaining them at zero over the entire simulation. Figure \ref{Fig5_quat_orient_vel_estimation} further demonstrates accurate estimations of the orientation and angular velocity. The robustness of the quaternion state representation were critical to the estimator's success during dynamic maneuvers. Unlike standard UKF applications that rely on Euler angles and risk gimbal lock singularities, the QUKF directly propagates unit-quaternions within the unscented transform. Throughout the interaction scenarios, the quaternion-based attitude representation maintained high numerical stability and strictly preserved the unit quaternion constraint via normalization at each update step. This fundamentally prevented long-term kinematic drift and provided a globally consistent, singularity-free estimation of the platform's pose under strong coupling between rotational and translational dynamics. It can be observed that the normalized QUKF angular velocity estimation error exhibits brief spikes during rapid attitude variations induced by external torques; however, these spikes remain bounded and quickly converge to zero. In contrast, the normalized estimation error signal of the Extended Kalman Filter (EKF) shows a delayed response accompanied by comparatively larger Root Mean Squared Error (RMSE). These results confirm that the QUKF provided the controller with highly accurate state estimates, ensuring responsive tracking of human operator guidance while effectively compensating for sensor noise.

\begin{figure*}[h!]
	\centering
	\includegraphics[width=1.1\textwidth]{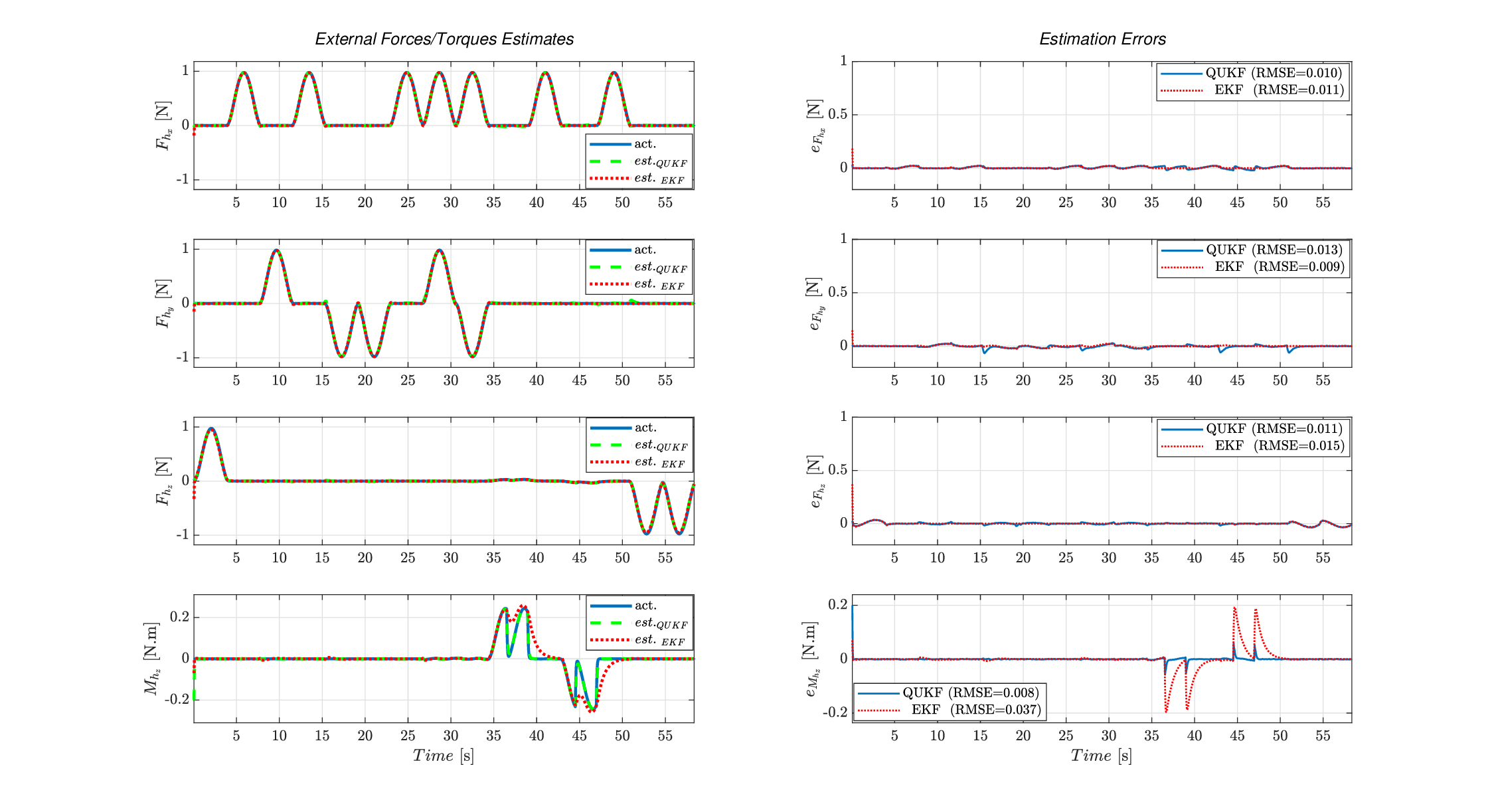}
	\caption{\label{Fig6:force_estimation}Actual vs. estimated external wrench with corresponding estimation errors: A performance comparison between the proposed QUKF and baseline EKF.}
\end{figure*}

The most critical evaluation is shown in Figure \ref{Fig6:force_estimation}, which compares the actual and estimated external forces and torque. The QUKF achieved near-zero estimation errors, demonstrating high precision in wrench estimation without requiring dedicated force–torque sensors. The method maintained a consistent estimation accuracy over more than 60 seconds of operation in the numerical simulation, validating its robustness for extended missions.
These results validate the effectiveness of the proposed approach
for human-UAV assistive payload transportation, offering a lightweight,
cost-effective alternative to systems that require dedicated force-torque
sensors while maintaining comparable performance in terms of interaction
quality and motion control.

To quantitatively validate the estimation accuracy, a comparative study was conducted against an Extended Kalman Filter (EKF). The quantitative results, summarized in Table \ref{tab:quantitative_comparison}, demonstrate that the QUKF consistently outperforms the EKF in terms of the Root Mean Squared Error (RMSE). Specifically, the QUKF achieved a substantial reduction in RMSE for angular velocities, with improvements of 43.65\% for roll rate ($p$), 66.41\% for pitch rate ($q$), and 59.25\% for yaw rate ($r$) compared to the baseline EKF. Furthermore, the QUKF demonstrated superior performance in wrench estimation, reducing the torque ($M_{h_z}$) estimation RMSE by 79.41\%. Additionally, the average convergence time of the estimation errors was highly responsive, recorded at ($t_c < 0.45$s), ensuring rapid tracking during human-UAV physical interaction. This significant improvement in accuracy confirms that the QUKF provides the superior robustness and precision necessary for reliable real-time state and wrench estimation in assistive human-UAV cooperative payload transportation.

\begin{table}[hbt]
	\centering
	{ \scriptsize
		\caption{Quantitative comparison of estimation accuracy (RMSE) between the baseline EKF and the proposed QUKF.}
		\label{tab:quantitative_comparison}
		\begin{tabular}{clccc}
			\hline 
			State & Metric & EKF (Baseline) & Proposed QUKF & \% Improvement\tabularnewline
			\hline 
			\hline 
			$x$ & RMSE ($m$) & 0.0040 & 0.00082 & $+79.50\%$ \tabularnewline
			$y$ & RMSE ($m$) & 0.0039 & 0.00083 & $+78.71\%$ \tabularnewline
			$z$ & RMSE ($m$) & 0.0040 &  0.00082 & $+79.50\%$ \tabularnewline 
			$\dot{x}$ & RMSE ($m/s$) & 0.0495 & 0.0404 & $+18.38\%$ \tabularnewline
			$\dot{y}$ & RMSE ($m/s$) & 0.0489 & 0.0393 & $+19.63\%$  \tabularnewline
			$\dot{z}$ & RMSE ($m/s$) & 0.0497 & 0.0396 & $+20.32\%$ \tabularnewline 
			$p$ & RMSE ($rad/s$) & 0.0181 & 0.0102 & $+43.65\%$ \tabularnewline
			$q$ & RMSE ($rad/s$) & 0.0652 & 0.0219 & $+66.41\%$ \tabularnewline
			$r$ & RMSE ($rad/s$) & 0.0265 & 0.0108 & $+59.25\%$ \tabularnewline 
			$F_{h_x}$ & RMSE ($N$) & 0.0105 & 0.0100& $+4.762\%$ \tabularnewline
			$F_{h_y}$ & RMSE ($N$) & 0.0091 & 0.0135 & $-48.35\%$ \tabularnewline
			$F_{h_z}$ & RMSE ($N$) & 0.0149 & 0.0109 & $+26.85\%$ \tabularnewline
			$M_{h_z}$ & RMSE ($N.m$) & 0.0374 & 0.0077 & $+79.41\%$ \tabularnewline
			\hline 
		\end{tabular}
	}
\end{table}

\section{Conclusion\label{Conclusion}}

In this work, an advanced Quaternion-based Unscented Kalman Filter (QUKF) was introduced for real-time estimation of system states and external interaction wrenches in assistive aerial payload transportation scenarios involving physical human-UAV interaction. By employing the unit quaternion formulation within the unscented filtering framework, the proposed method avoids singularities associated with Euler-angle representations and preserves the nonlinear structure of rotational dynamics, thereby enabling consistent and numerically stable states estimation. The effectiveness of this approach was quantitatively validated against the Extended Kalman Filter (EKF). Results show that the QUKF consistently achieves lower estimation errors, specifically reducing the RMSE of torque estimation by 79.41\% and improving the position and angular rate estimation by averages of over 79\% and 56\%, respectively. These findings highlight the potential of the QUKF as a promising strategy for aerial systems operating under physical interaction where precise state and wrench information is essential for feedback control and operational safety. While high-fidelity simulations underscore the potential of the proposed QUKF under controlled conditions, certain real-world factors, such as complex contact dynamics, payload impact responses, and unpredictable environmental disturbances, require further investigation. Consequently, experimental validation on a physical UAV-payload platform is necessary to fully assess robustness in realistic environments. Addressing these practical challenges through real-world experiments constitutes a primary direction for future research.

\section{Acknowledgment}

This work was supported in part by the National Sciences and Engineering
Research Council of Canada (NSERC) under the grants RGPIN-2022-04937.

\balance
\bibliographystyle{IEEEtran}
\bibliography{references2}

\end{document}